\documentclass{jfm}
\usepackage{graphicx}
\usepackage{setspace}
\usepackage{amsmath}
\usepackage{amssymb}
\usepackage{mathcomp}
\usepackage{textcomp}
\usepackage{xcolor}
\usepackage{siunitx}
\usepackage{etoolbox}
\usepackage{tikz}
\usepackage{array}
\usepackage[colorlinks,citecolor = blue, linkcolor=blue,hyperindex,CJKbookmarks]{hyperref}

\newcommand{\RomanNumeralCaps}[1]
\linenumbers
\newrobustcmd*{\mysquare}[1]{\tikz{\filldraw[draw=#1,fill=#1] (0,0)
rectangle (0.2cm,0.2cm);}}
\newrobustcmd*{\mycircle}[1]{\tikz{\filldraw[draw=#1,fill=#1] (0,0) circle [radius=0.1cm];}}
\newrobustcmd*{\mytriangle}[1]{\tikz{\filldraw[draw=#1,fill=#1] (0,0) --
(0.2cm,0) -- (0.1cm,0.2cm);}}
\usepackage[normalem]{ulem}

\shorttitle{Pinning induced motion and internal flow in neighbouring evaporating drops}
\shortauthor{P. J. Dekker, M. N. van der Linden and D. Lohse}

\title{Pinning induced motion and internal flow in neighbouring evaporating multicomponent drops}

\author{
    Pim J. Dekker\aff{1} \corresp{\email{p.j.dekker@utwente.nl}},
    Marjolein N. van der Linden \aff{2}\aff{1}
    \and Detlef Lohse\aff{1}\aff{3} \corresp{\email{d.lohse@utwente.nl}}
    }

\affiliation{\aff{1}Physics of Fluids group, Department of Science and Technology, Max Planck Center for Complex Fluid Dynamics and J. M. Burgers Centre for Fluid Dynamics, University of Twente, P.O. Box 217, 7500 AE Enschede, The Netherlands
\aff{2}Canon Production Printing Netherlands B.V., P.O. Box 101, 5900 MA Venlo, The Netherlands
\aff{3}Max Planck Institute for Dynamics and Self-Organization, Am Fassberg 17, 37077 G\"ottingen, Germany}

\begin{document}
\maketitle

\begin{abstract}
The evaporation of multicomponent sessile droplets is key in many physicochemical applications such as inkjet printing, spray cooling, and micro-fabrication. Past fundamental research has primarily concentrated on single drops, though in applications they are rarely isolated. Here, we experimentally explore the effect of neighbouring drops on the evaporation process, employing direct imaging, confocal microscopy, and PTV. Remarkably, the centres of the drops move away from each other rather than towards each other, as we would expect due to the shielding effect at the side of the neighbouring drop and the resulting reduced evaporation on that side. We find that pinning-induced motion mediated by suspended particles in the droplets (due to contamination or added on purpose) is the cause of this counter-intuitive behaviour. 
With the help of direct numerical simulations we also explore the relative contributions of the replenishing  flow and of the solutal and thermal Marangoni flows to the overall flow dynamics in the droplet. Finally, the azimuthal dependence of the radial velocity in the drop is compared to the evaporative flux and a perfect agreement is found.

\end{abstract}


\section{Introduction} \label{sec:intro}
The evaporation of sessile droplets has received considerable attention, both because of fundamental scientific interest and because of its importance to many technological and biological applications, such as inkjet printing \citep{hoath2016,wijshoff2018,lohse2022}, spray cooling \citep{cheng2016}, nanofabrication \citep{brinker1999}, pesticide spraying \citep{yu2009,gimenes2013}, and diagnostics \citep{zang2019}. The interest in evaporating drops has taken a leap since the description of the “coffee stain effect” by \citet{deegan1997}. Since then, major progress has been made in the understanding of evaporating drops \citep{zang2019,lohse2022,gelderblom2022,wang2022,wilson2023}.

The flow in isolated evaporating drops has been extensively studied \citep{deegan2000,larson2014,edwards2018,diddens2021,gelderblom2022}. For small drops, i.e. smaller than the capillary length, $l_c=(\gamma/\rho g)^{1/2}$, the drop shape is a spherical cap \citep{degennes1985}. In an evaporating drop, a replenishing (capillary) flow must develop for a drop to keep this shape. For given volatility and given other material properties of the liquid, the drop lifetime also depends on the behaviour of the contact line \citep{nguyen2012ces,stauber2014,stauber2015} and on evaporative cooling \citep{ristenpart2007,dunn2009}, i.e., on the substrate properties. 

Additionally, there can be surface tension gradients on the drop surface which lead to a Marangoni flow. These surface tension gradients can arise due to evaporative cooling, due to compositional gradients of the drop constituents because of selective evaporation, or because of the presence of surfactants at the interface. Typically, the thermal Marangoni flow due to evaporative cooling is much weaker as compared to the solutal Marangoni flow \citep{gelderblom2022}. When the Marangoni flow (either solutal or thermal) in the drop is sufficiently large, the apparent contact angle of the drop can increase, which is known as Marangoni contraction \citep{karpitschka2017,shiri2021,kant2024}. In the extreme case, the shape can even deviate from a spherical cap shape \citep{pahlavan2021}.

In general, not only surface tension can depend on the composition of binary or multicomponent drops, but many other physical properties also depend on the composition, such as vapour pressure, density, surface tension, viscosity, and the diffusion coefficient, giving rise to a huge parameter space. Moreover, for mixtures consisting of three or more components, the diffusion coefficient becomes a matrix with $(n-1)^2$ elements \citep{legros2015}. The off-diagonal terms are usually small \citep{cussler2009}. 

\begin{figure}
    \centering
    \includegraphics[width=1\textwidth]{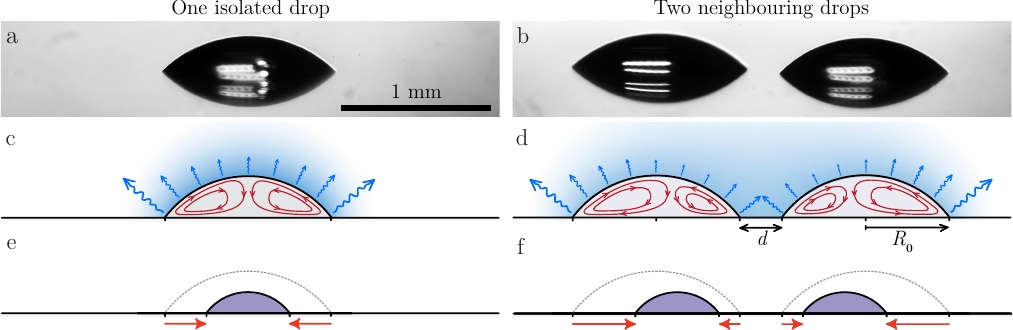}
    \caption{
        Evaporation of an isolated drop (left column) and two neighbouring drops (right column). (a,b) Experimental snapshot shortly after deposition. Scale bar applies to both images. (c) The arrows in the drop indicate the Marangoni flow in the drop. The arrows on the drop surface indicate the evaporative mass flux. The colour gradient indicates the increased relative humidity around the drop. (d) Two identical drops with radius $R_0$ that are placed at a distance $d$ apart. In the central region between the drops, the evaporation is reduced due to the shielding effect. (e/f) the \textit{expected} final positions of the drop after all water has evaporated. The grey dashed line shows the original drop position. (e) Assuming the drop is initially unpinned the contact line will recede on all sides equally. (f) Since less fluid evaporates in the region close to the neighbouring drop, the contact line will recede less as compared to the regions far from the neighbouring drop. As a result the drop centres have moved closer together.
    }
    \label{fig:side}
\end{figure}

Recently, due to the relevance for applications, more attention has also been given to the evaporation of multiple neighbouring droplets. In fig.\ \ref{fig:side} we contrast the situation for an isolated droplet (a,c,e) to that of two neighbouring droplets (b,d,f). When the  isolated drop of water evaporates, the surrounding water vapour concentration is increased. When the second drop is close enough, the water vapour concentration will be further enhanced. This locally increases the ambient relative humidity that each drop experiences compared to the case of an isolated drop, which in turn increases the lifetime of neighbouring drops \citep{fabrikant1985,laghezza2016,hatte2019,khilifi2019,chong2020,schofield2020,edwards2021,masoud2021}. Additionally, the evaporative flux $J = J(r,\theta)$ on the surface of each drop is no longer axisymmetric but is now a function of the azimuthal angle \citep{wray2020}. For example, for two drops close to each other, the side of the drop that is facing the neighbouring drop is shielded more than the part of a drop that faces away from the neighbouring drop. This means that the evaporative flux in the region between the drops is reduced by the presence of the other drop, as illustrated in figure \ref{fig:side} (f). The expected consequence thereof is that the centers of mass of the two droplets move towards each other, as there is relative more evaporation at the side of the droplet pointing away from the other drop. However,  in our experiments we will observe  the {\it opposite} behaviour (see fig.\ \ref{fig:snapshots}): The centers of mass of the droplets move away from each other! We found this behaviour remarkable, surprising, and counter-intuitive. Why do the two drops behave like that and can one understand the physics of this behaviour? 

The answer to this question partially lies in the asymmetric flux in the droplets which affects the solute deposition around the drop and therefore the pinning behaviour. The amount of solute deposited will be smaller in regions where the flux is reduced the most and is larger where the flux is reduced the least \citep{wray2021}. Additionally, the solutal Marangoni flow will also become asymmetric. When the two drops also experience Marangoni contraction, the drops will start to move since the Marangoni contraction is reduced on the side of the other drop. Generalising to beyond the system that is studied in this work, the drops can attract, repel, and even chase each other, depending on the composition and volatility \citep{cira2015}.

In addition to answering the above question, the more general objective of the paper is to explore the internal flow in evaporating multicomponent neighbouring droplets and in particular elucidate the role of contact line pinning in the flow dynamics. Both are crucial for many applications where either the position of the drop or the deposition of solute is important, with inkjet printing as prime example \citep{lohse2022}, but in spite of the prevalence of  such multicomponent neighbouring droplets  in nature and technology, little is known about the flows inside such drops and how they affect each other. We will focus on binary drops of water with 1,2-hexanediol, which is non-volatile and has a lower surface tension than water, resulting in solutal Marangoni flow in evaporating water--1,2-hexanediol droplets. It is this property which makes such droplets an excellent and highly relevant model system for aqueous inks in inkjet printing. We will investigate the flow in these evaporating neighbouring multicomponent drops  using confocal microscopy for PTV and synchronized side view shadowgraphy in a humidity-controlled chamber. We will show that this flow is indeed asymmetric as expected, and we find perfect agreement with the local evaporative flux. 

The paper is organized as follows: In section 2 we describe our experimental methods. In section 3 we report the observed contact line behaviour. The core of the paper is section 4, in which we report and explain the observed internal flow inside the droplets, whose central contributions are the pinning-induced replenishing flow and the Marangoni flow. The relative importance of solutal and thermal Marangoni flow is  elucidated in section 5, with the help of direct numerical simulations, in which it is easier to turn off either of them. That section is supplemented by Appendix A, in which we further discuss and rationalize why the effect of thermal Marangoni flow is so limited. The paper ends with a summary,  conclusions, and an outlook (section 6).


\section{Experimental procedure}  \label{sec:methods}
The experiments were performed in a humidity and temperature-controlled chamber. During the measurements the airflow was disabled. The chamber volume $\approx 0.75$ l was sufficiently large to ensure diffusion-limited evaporation. The binary mixtures consisted of 10 wt$\%$ 1,2-hexanediol (Sigma-Aldrich, 98$\%$ pure) which is non-volatile, and 90 wt$\%$ water (“Milli-Q”, resistivity 18 M$\tcohm$cm) which is volatile. The drops were deposited on a glass slide coated with octadecyltrichlorosilane (OTS) which is a transparent hydrophobic coating \citep{silberzan1991}. The advancing and receding contact angle of a water drop on this substrate are $\theta_\mathrm{advancing}=110.3^\circ \pm 0.6^\circ$ and $\theta_\mathrm{receding} = 89.7^\circ \pm 1.1^\circ$. A detailed procedure of the substrate preparation and the contact angle measurements are provided in the SI.

Both drops were imaged from the side using a camera (Nikon D850) and a long-distance microscope at 7x magnification (Navitar 12x Zoom). One of the drops was simultaneously imaged from below with a confocal microscope (Nikon A1R). We were able to measure the flow in the drop by adding a small amount of polystyrene fluorescent particles (microParticles GmbH, concentration = $3.8\cdot10^{-3}$ vol$\%$, diameter $= 1.14\,\tcmu$m, $\lambda_{absorption} = 530$ nm, $\lambda_{emission} = 607$ nm) and tracking these particles with a particle tracking velocimetry (PTV) algorithm. Since the confocal microscope has a very narrow depth of field, we know exactly at what height in the drop we detect the particles. We chose to image close substrate ($\approx 18 \ \tcmu$m) such that the drop remains in focus as it evaporates. The confocal microscope is able to provide a bright field view of the drop (Transmission Detection) at the same time as the fluorescent image. More details of the experimental setup and data analysis are provided in the SI.

Due to the constraints of the confocal and the side view camera in combination with the humidity chamber, it was not feasible to implement an automatic dispensing system for the drops. Therefore all drops were deposited manually deposited using a syringe (Hamilton, model 7001) capable of accurately expelling very small volumes (syringe volume = 1$\tcmu$l). The volume of each drop was $0.15 \, \tcmu \mathrm{l}$, which is the largest drop size that fits completely in the field of view of the confocal microscope. The time between the depositing of the first drop and the second drop (10 s) was much shorter than the total lifetime of the drops (4.5 minutes). Although precise repetitions of drop placement were not possible, the distance between the drops could be accurately determined using the side view camera in combination with the top view of the confocal after each experiment.


\section{Contact line behaviour} \label{sec:pinning}

\begin{figure}
    \centering
    \includegraphics[width=0.55\textwidth]{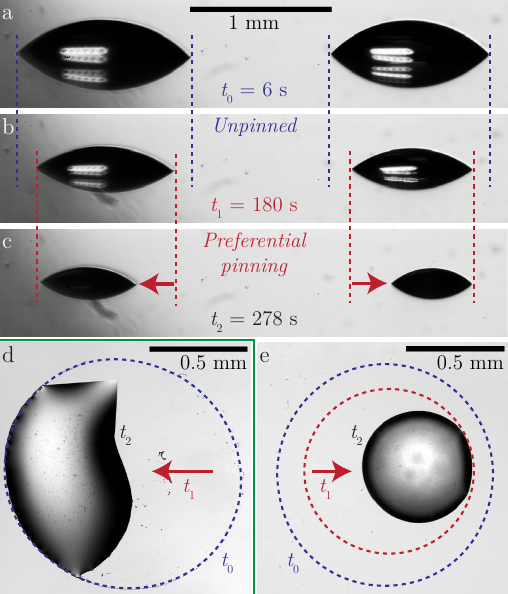}
    \caption{(a,b,c) Show three snapshots of the side view of two drops evaporating. Image (a), taken at $t_1=6$ s, shows how the two drops are deposited. The position of the contact line is marked with the dashed line and is extended into the next snapshot for reference. For the first 180 s of the evaporation, the drops are unpinned. In image (b), taken at $t_1=180$ s, both drops get pinned only on the far side w.r.t. the other drop. Because of the partial pinning of the drops they start to move apart. In image (c), taken at $t_2 = 278$ s, the water has stopped evaporating. Image (e) shows the bottom view of the right drop shown in (a,b,c). The dashed lines correspond to $t_0$, the original size of the drop, and $t_1$, the moment the drop becomes partially pinned. The left drop looks correspondingly but moves to the other side. Image (d) shows a different case where the drop was pinned from the start but unpinned at some time on one side only. The other drop was on the right of the drop in (d).}
    \label{fig:snapshots}
\end{figure}

\begin{figure}
    \centering
    \includegraphics[width=0.9\textwidth]{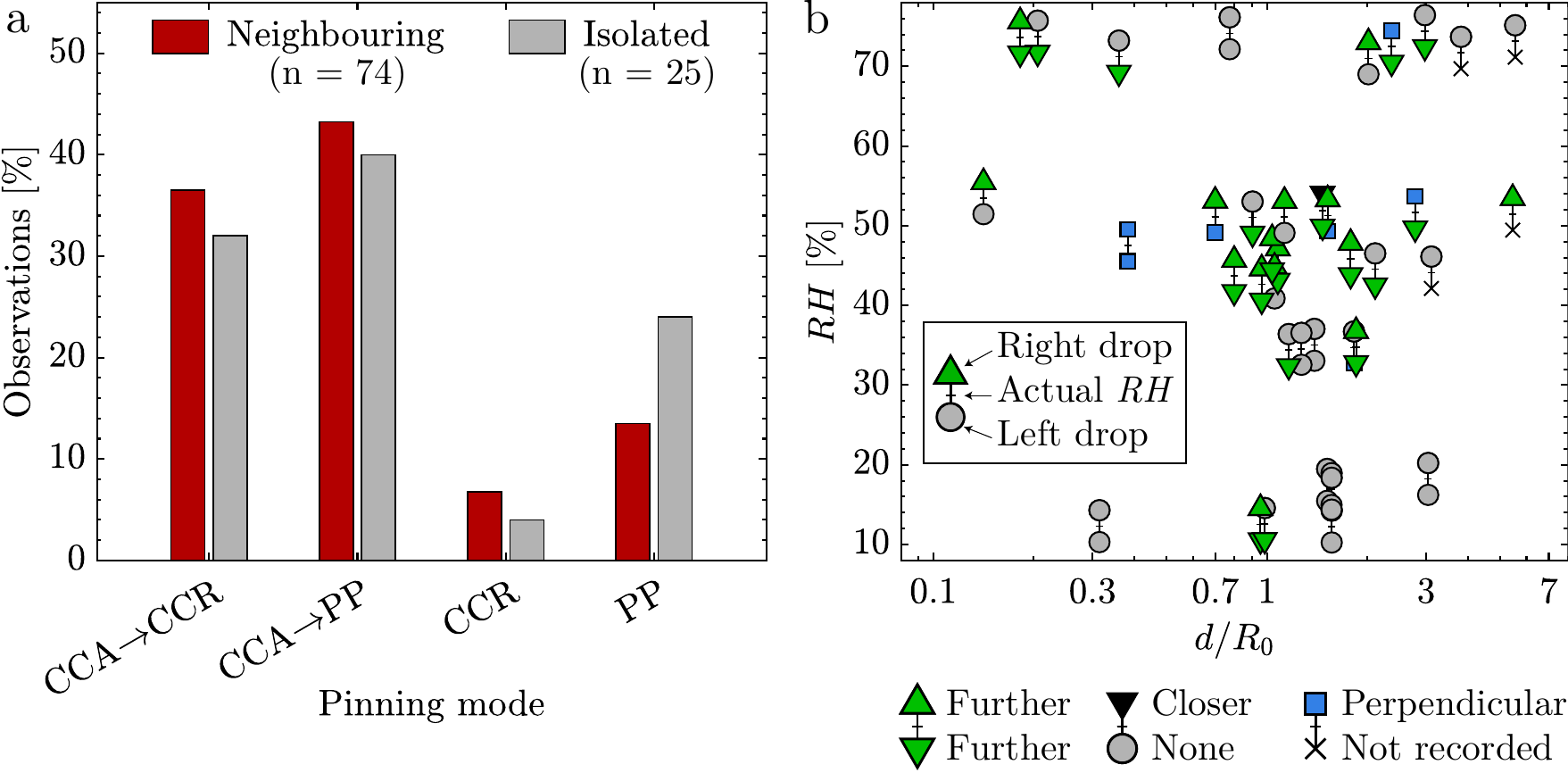}%
    \caption{(a) The frequency different contact line behaviours were observed for 74 neighbouring drops (dark re bars) and 25 isolated drops (light grey bars). The different pinning modes are: CCA (constant contact angle, i.e. the contact line recedes on all sides of the drop), CCR (constant contact radius, i.e. the contact line is pinned on all sides of the drop), and PP (partial pinning, i.e. part of the contact line is pinned and another part of the contact line is receding). (b) Direction of movement of neighbouring drops for different initial separations and relative humidities. Since each pair of drops would overlap, we split the drops vertically around the actual relative humidity (indicated with a small bar). The drop movement is shown with a triangle pointing in the direction of the drop movement with respect to the other drop. A square is shown when a drop did move, but did not get closer or further w.r.t. the other drop. Drops that did not move are circles. When a drop was not in view an $\times$ is shown.}
    \label{fig:pinning_stats}
\end{figure}

Figure \ref{fig:snapshots} (a,b,c) shows snapshots of the side view at different times of two evaporating drops. Figure \ref{fig:snapshots} (e) shows the bottom view of the right drop in figure \ref{fig:snapshots} (c). Figure \ref{fig:snapshots} (a) shows the drops just after they have been deposited on the substrate. The initial drop contact radius is $R_0 = 0.53$ mm and the distance between the drop edges is $d/R_0 = 1.7$. As a visual guide, the position of the contact line has been marked by a dashed line, which extends into the next panel. The position of the contact line is also marked in the bottom view with a dashed line in figure \ref{fig:snapshots} (e). For the first 180 s after the drop deposit, the drop evaporates in a constant contact angle mode (CCA). The contact line of each drop moves the same distance inwards on all sides. Up to that time, the centres of both drops remain in the same place. 

However, after some more time, we see that the centres of the drops move \textit{away} from each other. Figure \ref{fig:snapshots} (b) shows the moment when the drops start to move away from each other. Again the position of the contact line is marked with a dashed line. Figure \ref{fig:snapshots} (c) shows the residue of the non-volatile 1,2-hexanediol when the evaporation of water in the drop has stopped. Going from panel (b) to (c), we see that the contact line facing away from the other drop has stayed in place, while the contact line facing toward the other drop has retracted. We will call this mode partial pinning (PP). During the whole evaporation, the contact angle of both drops remained the same.

In our experiments, this was not the only mode of evaporation that we observed. Figure \ref{fig:snapshots} (d) shows a different case of pinning-induced motion (the left drop is shown, with the other drop being on the right). Here, the drop was pinned from the start and evaporated in constant contact radius mode (CCR). At time $t_1=102$ s the drop suddenly unpinned on one side and moved away from the other drop. This is different for the drops shown in (a,b,c), which were evaporating in CCA mode and got pinned on one side during the evaporation. In both cases, figure \ref{fig:snapshots} (d) and (e), the result is that the drops move apart. 

In total we observed four different contact line behaviours. Figure \ref{fig:pinning_stats}(a) shows the frequency of each observed mode for all the experiments. CCR: In this cases the drop would not exhibit any pinning-induced motion and evaporate in CCR mode during its entire lifetime. PP: In this case the drop is partially pinned, meaning that a section of the contact line is pinned and does not move, whereas the rest of the contact line recedes inwards, resulting in a shift of the drop centre. In the majority of cases (78$\%$), both the CCR and PP modes can be preceded by a period of time where the drops evaporates in CCA mode. Meaning that the contact line recedes on all sides of the drop equally. We denote these modes as $\mathrm{CCA \rightarrow CCR}$, and $\mathrm{CCA \rightarrow PP}$.

The drop(s) moved due to partial pinning in 59$\%$ of cases. Isolated drops also showed partial pinning, although the direction in which the centre moved was arbitrary. For neighbouring drops, however, the direction was systematically away from the other drop. Figure \ref{fig:pinning_stats}(b) shows the direction in which the centres of the drops moved relative to the other drop for different relative humidities and drop separations. When the drops moved, 80$\%$ of the cases they moved further apart. Sometimes the drops would move, but they would not get any closer of further apart, i.e., they moved perpendicular with respect to each other. This happened 17$\%$ of the cases. In one experiment, one of the drops moved toward to the other drop, while the other drop moved away from the other drop. For the entire range of relative humidities and drop separations, both the (CCA$\rightarrow$) PP modes as well as the (CCA$\rightarrow$) CCR modes can be observed. Indicating there is no clear dependence on the relative humidity or drop separation.

We emphasise that the movement of the drops centres \textit{away} from each other is surprising: the direction is opposite to what would be predicted based on the shielding effect. Since the local evaporation rate of the drop is suppressed near the other drop, we would expect the contact line to recede less, since less mass is lost. And because the evaporation is highest in regions facing away from the other drop, we would expect the contact line to recede more quickly in those regions. Thus, we would expect the centre of the drop should move towards the other drop. 

When we consider the effect of the flow on the contact line motion we also expect the opposite behaviour: The flow inside an evaporating water/1,2-hexanediol drop is dominated by a Marangoni flow (as will be discussed in section \ref{sec:ptv} and \ref{sec:num}) which can result in Marangoni contraction \citep{hack2021}. When the two drops are close, the Marangoni flow becomes asymmetric due to the shielding effect and cause the drops to contract more where the evaporation rate is higher. Consequently we expect the centres of the drops to move closer \citep{cira2015}. However, we do not observe any Marangoni contraction --- during the first part of the evaporation process ($t_0 \rightarrow t_1$) we observe no significant displacement of the drops. Marangoni contraction depends strongly on the wettability of the substrate. Both \citet{hack2021} and \citet{cira2015} used aggressive cleaning methods to make the glass perfectly clean. Which is very different from the glass substrates coated with OTS that we study.

\begin{figure}
    \centering
    \includegraphics[width=0.5\textwidth]{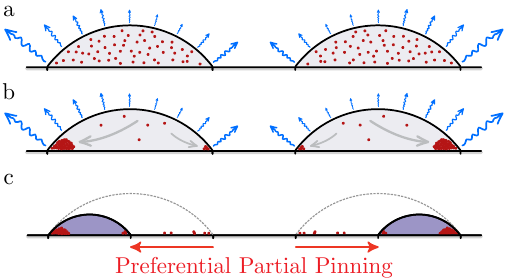}%
    \caption{Schematic of the preferential partial pinning effect. (a) Initially, the particles are distributed homogeneously throughout the drop. Due to the shielding effect, evaporation is suppressed in the area between the drops. (b) During evaporation this will result in most particles agglomerating at the side opposite the other drop due to the higher flux there. This will make it more likely that the drop pins on that side only because of the larger number of particles there. (c) The \textit{observed} final position of the drop. The dashed line is the initial position of the drop.}
    \label{fig:schematic}
\end{figure}

We rationalise the observed pinning induced motion by considering that the motion of the drops is strongly influenced by particles and solutes present in the drop. Due to the strong evaporation at the rim of the drop, particles or solute will accumulate at the contact line. Despite the Marangoni flow mixing the particles throughout the drop \citep{raju2022}, more particles will deposit and accumulate at the contact line where the evaporative flux is the largest \citep{wray2021}. This results in a large concentration of particles in the region facing away from the other drop, increasing the likelihood of the drop pinning there, as is illustrated in figure \ref{fig:schematic}. 

So far, we only have considered a drop of water/hexanediol with a small amount of tracer particles to facilitate the PTV. To investigate the role of the tracer particles we repeated the experiment without tracer particles. Of all the experiments with neighbouring drops (n = 14), 11 (79$\%$) moved due to partial pinning. In 9 cases the drop centre moved further away from the other drop (82$\%$). Although the number of the experiments without tracer particles is limited, we still observe the pinning-induced motion. This means that the added tracer particles contribute little to the contact line behaviour. This observation does not invalidate our hypothesis, since we still observe very small particles or contaminants in the drop with the microscope at very high magnification. We speculate that the largest contributor of particles is from the surrounding air, which contains particulate matter of various sources (much) smaller than $1 \ \mathrm{\tcmu m}$ such as dust, soot, and aerosols \citep{zhang2015}. Future experiments in a clean room might disentangle the role of particulate matter on the contact line dynamics. However, these contaminants are simply unavoidable and uncontrollable in a standard laboratory setting or in most technological applications.

\begin{figure}
    \centering
    \includegraphics[width=.55 \textwidth]{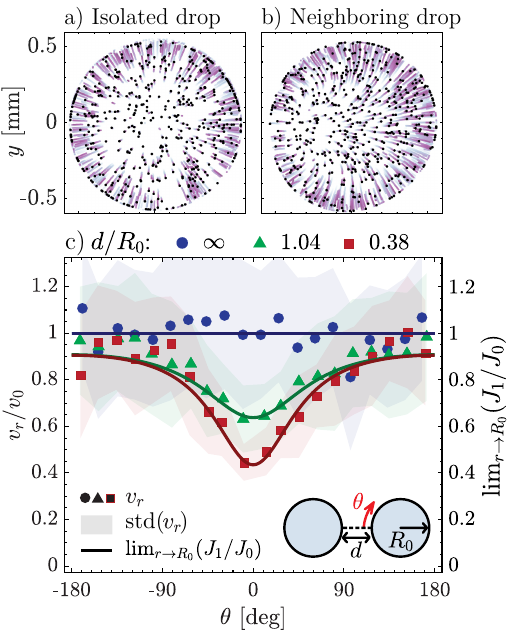}
    \caption{Flow in $\mathrm{RH} = (47 \pm 5) \, \%$ and $T = (23.8 \pm 0.3) \,\tccelsius$. (a,b) Show the traces of particles in the drop close to ($\approx 18\  \tcmu$m) the substrate for the first 15$\%$ of the total evaporation time. (a) Is for an isolated drop. (b) is for a neighbouring drop with $d/R_0=0.38$. (c) The azimuthal dependence of the averaged radial velocity in the drop close to the rim is shown for three drop separations: $d/R_0 = \{ \infty, \, 1.04, \, 0.38 \}$ with the symbols $\{$\mycircle{black}, \mytriangle{black}, \mysquare{black}$\}$, respectively. The other drop is located at $\theta = 0$. The shaded area is the standard deviation of the average velocity distribution. The solid lines are the azimuthal dependencies of the flux for the neighbouring drops normalized by the flux of an isolated drop: $\displaystyle{\lim_{r \to R_0}{J_1/J_0}}$.}
    \label{fig:ptv}
\end{figure}

\section{Internal flow} \label{sec:ptv}
To further elucidate the interplay between pinning, evaporation, and Marangoni flow, we have imaged the flow in the drops with PTV (Particle Tracking Velocimetry). Figure \ref{fig:ptv} (a) and (b) show the trajectories of the particles close to the substrate during the first 15$\%$ of the total evaporation lifetime. By limiting the analysis to a short time after the drop deposition, we can ignore any effect of the contact line motion, as well as the unusual segregation dynamics that occurs in the final phase of evaporation of water/1,2-hexanediol drops \citep{li2018}. Figure \ref{fig:ptv} (a) shows the trajectories for an isolated drop. The flow is outwards and axisymmetric. There is a stagnation point in the centre of the drop where the particles do not move. Near the rim of the drop, the particles move the most. Some particles are stuck at the contact line.

Figure \ref{fig:ptv} (b) shows the trajectories for a neighbouring drop (with the neighbouring drop on the left of the drop that is shown). The flow is still outwards, but no longer axisymmetric. The stagnation point has shifted towards the other drop. We also see that the displacement of particles is much larger on the right (away from the other drop) than on the left. The evaporative flux is larger on the right of the drop as the left side is shielded by the other drop. More water evaporates on the right leading to larger concentration gradients in the drop, and consequently larger surface tension gradients. This results in a stronger Marangoni flow, which dominates a larger portion of the drop, shifting the stagnation point towards the other drop.

To better quantify the asymmetry in the velocity in the neighbouring drop, we compute the average velocity of the particles close to the rim for the first 15$\%$ of the drop lifetime. This time interval was chosen such that a sufficient large number of particles were measured, while limiting the effects of volume loss of the drop and of the contact line motion on the measurement. We include all the particles on the interval $r/R_0 = \left[0.8, 0.85\right]$. When we choose an interval closer to the rim the results become unreliable due to the limited depth of field of the confocal microscope, which brings the flow near the water-vapour interface in focus. Choosing a smaller space and time interval is possible but will result in more noise and a larger statistical error in the data. Figure \ref{fig:ptv} shows the average radial velocity close to the rim as a function of the azimuthal angle $\theta$ in the drop for different drop separations, normalized by the average velocity of an isolated drop. The shaded region corresponds to the standard deviation for each bin.

For $d/R_0 = \infty$ the average velocity is axisymmetric (i.e. constant for all $\theta$). For $d/R_0 = 1.04$, the overall velocity is slightly lower, and a clear minimum appears around $\theta = 0$. For $d/R_0 = 0.38$, the minimum becomes more pronounced, confirming the observations made in figure \ref{fig:ptv} (b).

Next, we also would like to make a quantitative comparison with the local evaporation rate. To this end we will have to make a few simplifications, since to our knowledge, there is no expression of the local evaporative flux for multicomponent drops. However, the flux has been investigated for single-component drops with a contact angle of zero, i.e. flat disks. \citet{fabrikant1985} analysed the potential flow through membranes, which is mathematically equivalent to neighbouring evaporating drops. Later, \citet{wray2020} continued the analysis and calculated the local evaporative flux $J_1$ of two identical neighbouring drops, 

\begin{align}
    J_1(r,\theta) = J_0(r)f(r,\theta).
   \label{eq:J1}
\end{align}

\noindent Here $J_0$ is the evaporative flux for an isolated drop,

\begin{align}
   J_0(r) = \frac{2 D_{vap} \Delta c_{vap}}{\pi \sqrt{R^2 - r^2}},
   \label{eq:J0}
\end{align}

\noindent and $f(r,\theta)$ expresses the shielding effect due to the other drop \citep{wray2020},

\begin{align}
   f(r,\theta) = 1-\frac{F \sqrt{b^2-R^2}}{2 \pi \left(r^2 + b^2 - 2 r b \cos\theta\right)}.
    \label{eq:f}
\end{align}

\begin{figure}
    \centering
    \includegraphics[width=0.70\textwidth]{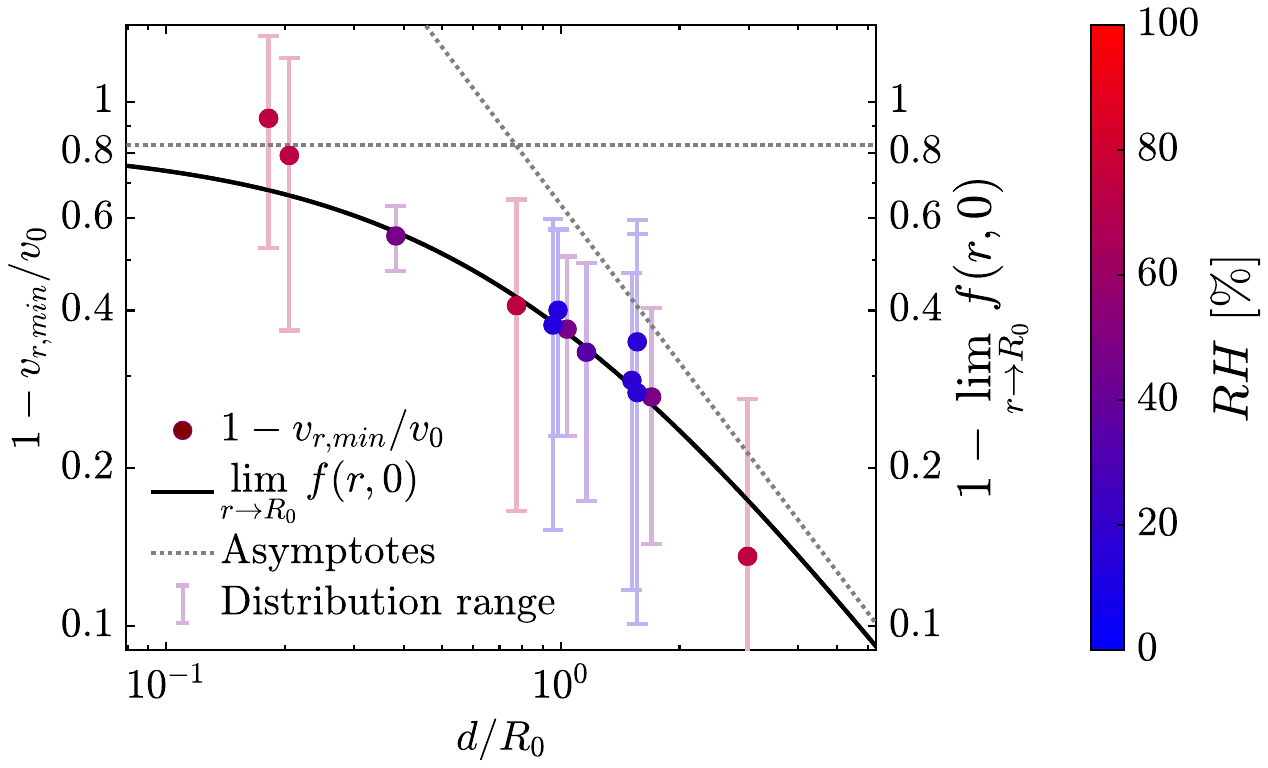}%
    \caption{The minima in the radial velocity measured between $0.8<r/R_0<0.85$ for the first 15$\%$ of the drop lifetime normalized by the value for an isolated drop, for various drop separations and relative humidities. The relative humidity is color coded using the color scale shown on the side. The bars around each datapoint correspond to the standard deviation of the distribution of all velocities included in the average. The solid line is the minimum in the local evaporative flux close to the rim, normalized by an isolated drop, as a function of the drop separation (equation \ref{eq:f}). The asymptotic limits of the flux are shown by the dashed lines: $4/(3\sqrt{3})$ for $d/R_0 \rightarrow 0$, and $2R_0/(\pi d)$ for $d/R_0 \rightarrow \infty$.}
    \label{fig:dimple}
\end{figure}

\noindent The drop separation is given by $b$ which is the distance from the centres of each drop ($b = d+ 2 R_0$), where $f(r,\theta)$ is the local shielding, and $F$ is the overall shielding of the flux, given by $F = 4\pi R/(\pi+2\arcsin{(R/b)})$. Note that $R_0$ is the initial contact radius and that $R$ depends on time (i.e. $R_0 = R(t=0)$). Even though equations \ref{eq:J0}-\ref{eq:f} are only valid for single component drops, this can still be a good approximation for the early stages of the evaporation since the drop mostly consists of water with only relatively small concentration differences throughout the drop.

Since the flow in the drop is driven by the evaporative flux, we can try to directly compare the measured radial velocity to the result of \citet{wray2020}. The solid line in figure \ref{fig:ptv} is the evaporative flux close to the rim for a neighbouring drop, normalized by an isolated drop, i.e. the shielding of the flux by the neighbouring drop: $f(r,\theta)$ evaluated at $r\rightarrow R_0$. Note that no fitting parameters were used. The shape of the curve only depends on the separation between the drops, which was measured using the side view camera.

The agreement between the local flux and the radial velocity is remarkably good. We would like to emphasize that this relationship is not trivial, since we observe a strong Marangoni flow in the drop, which indirectly depends on the evaporative flux. The shielded evaporative flux in combination with selective evaporation leads to the concentration gradients in the drop. This in turn leads to surface tension gradients which then drives the Marangoni flow in the drop. It is very reasonable to expect that the flux and radial velocity share \textit{qualitatively} the same shape, but here the agreement is also \textit{quantitative}. This suggests that the relation with the local flux and velocity close to the rim is directly proportional to each other: $J\sim v_r$.

To further substantiate the relation between the flux and the radial velocity, we investigate the minimum of $J$ and $v_r$ as shown in figure \ref{fig:ptv}(c). Figure \ref{fig:dimple} shows $J$ and $v_r$ at $\theta = 0$ as a function of the drop separation for various relative humidities. The solid line is the minimum in the local evaporative flux for neighbouring flat drops. The discontinuous lines indicate the asymptotic limits for small and large drop separations. The data points are the minima in the measured average radial velocity in neighbouring drops, the color indicates the relative humidity. The bars indicate the spread of the distribution using the standard deviation of all measured velocities included in the average.

Despite the large spread of the distribution for the measured radial velocity, the agreement between the velocity and the flux is very good. For very small drop separations we find that the velocity deviates from the flux. This is because we are limited in how near we can measure to the contact line. For very large drop separations we also observe a deviation. This is because the magnitude of the fluctuations in the average radial velocity are similar to the minimum in the radial velocity. For all intermediate drop separations the minimum in the flux and velocity match.


\section{Direct numerical simulations on the relative importance of solutal and thermal Marangoni flow} \label{sec:num}

We now want to further elucidate the role of solutal and thermal Marangoni force on the flow in the drop. We will do so by direct numerical simulations, as it is then possible to fully turn off either the solutal Marangoni forces, or the thermal ones, or even both. Unfortunately, it is not easily feasible to directly simulate two neighbouring droplets -- at least not without major simplification -- since it is numerically impossible to fully resolve the required  3D mesh due to the prohibitively large computation cost this would require. Therefore in this section we will focus on isolated droplets only, which we can numerically resolve by making use of the axisymmetry. For evaporating isolated water--1,2-hexanediol droplets this axisymmetry is given during most of the evaporation time; only in the final phase it may be broken as then the surface tension of the water--1,2-hexanediol mixture non-monotonously depends on the 1,2-hexanediol concentration, cf.\ \cite{diddens2024}. In this section we  will focus on the effect thermal Marangoni has when solutal effects are present. The case of purely thermal Marangoni flow is less relevant for our experiments. We will therefore discuss it only in appendix \ref{app:num_th_only}. For that case, we will find that the flow pattern is quite different. 

\subsection{Numerical Methods}

\begin{figure}
    \centering
    \includegraphics[width=0.85\textwidth]{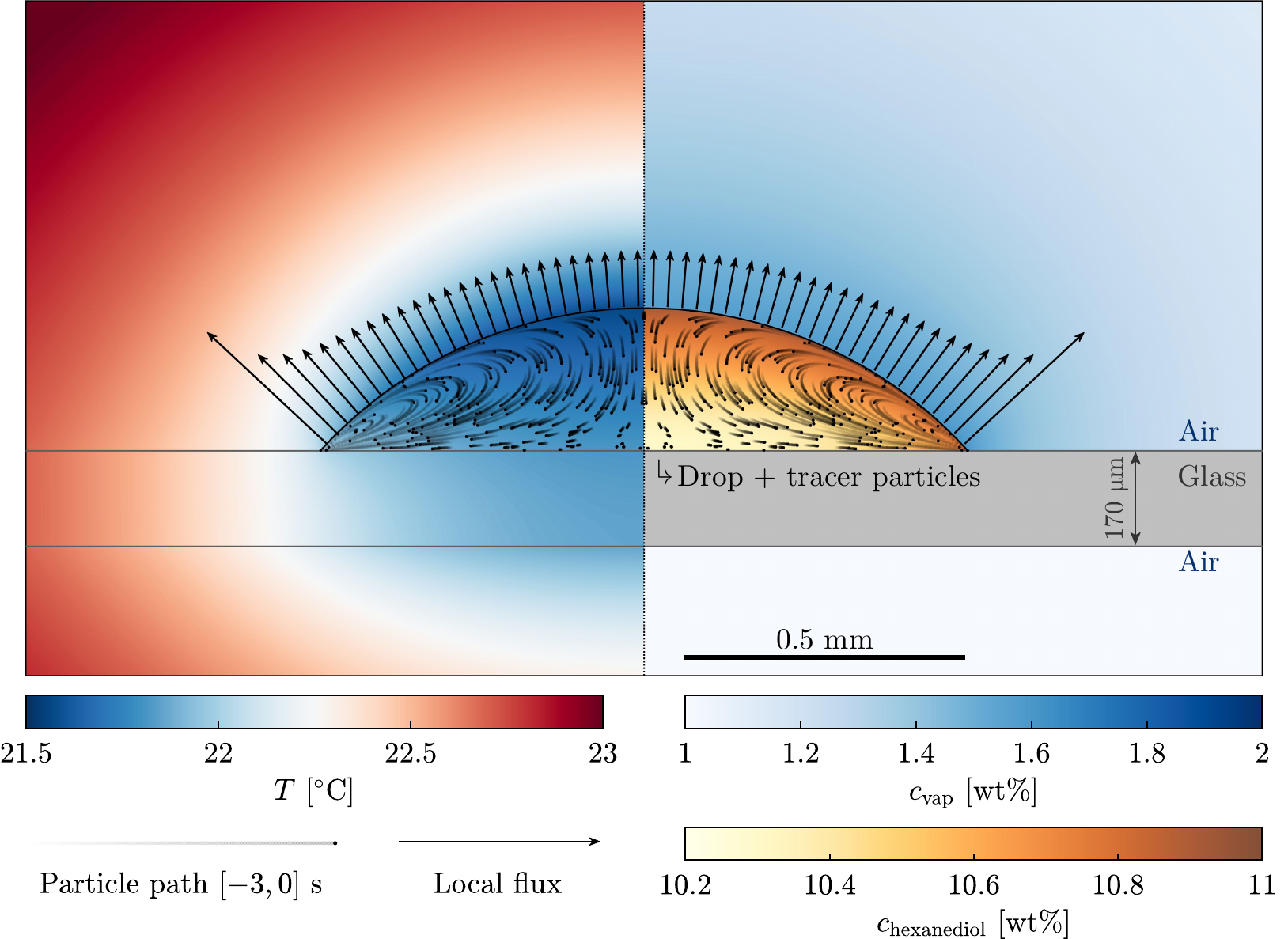}%
    \caption{Snapshot at $t$ = 15 s of a numerical simulation of an isolated water/hexanediol drop evaporating on a glass substrate in the same conditions as the experiment shown in figure \ref{fig:ptv} ($c_{0,\mathrm{hexanediol}} = 10 \,\mathrm{wt\%},\ RH = 50\%,\ T = 24\mathrm{^\circ C},\ \theta = 50^\circ, \ V = 0.15\,\mathrm{\tcmu l}$). The simulations are axisymmetric and include thermal effects. The left half shows the temperature and the right half shows the concentration of either water vapour in the gas phase of hexanediol in the drop. The arrows indicate the local evaporative flux and the trajectories of tracer particles show the flow in the drop.}
    \label{fig:num_overview}
\end{figure}

Numerical simulations are done using pyoomph (source code: \url{https://github.com/pyoomph/pyoomph}, documentation: \url{https://pyoomph.readthedocs.io}), which is a python multi-physics finite element framework based on oomph-lib (\url{http://www.oomph-lib.org/}) and GiNaC (\url{http://www.ginac.de/}). A detailed description of the used model can be found in \citet{diddens2017jcs}.
\\ \\
Figure \ref{fig:num_overview} shows a numerical snapshot after 15 seconds of evaporation. Similar to \citet{diddens2017} we simulate the whole domain (gas, substrate, and liquid) with thermal effects enabled, as shown on the left side of figure \ref{fig:num_overview}. The substrate is modelled after the glass slides used in the experiments and has the thermal properties of glass and a thickness of 170 $\tcmu$m. The ambient temperature is set to $24\mathrm{^\circ C}$, the relative humidity to $RH = 50\%$, mimicking the conditions of the experiment shown in figure \ref{fig:ptv}. The right side of figure \ref{fig:num_overview} shows the concentration of water vapour in the gas phase. The only method of water vapour transfer is diffusion and natural convection is ignored. The arrows indicate the local evaporative flux of water into the gas phase. The drop itself is pinned (i.e. CCR mode). Inside the drop we consider both convection and diffusion of hexanediol using physical properties of water/1,2-hexanediol \citep{li2018}. We add tracer particles with a trail to visualise the flow inside the drop. The particles are only visual and are not coupled to the flow. We can study the effect of solutal and thermal Marangoni flow by making the surface tension independent of composition and/or temperature at the liquid-gas interface of the drop.

\subsection{Results from the direct numerical simulations}

\begin{figure}
    \centering
    \includegraphics[width=0.80\textwidth]{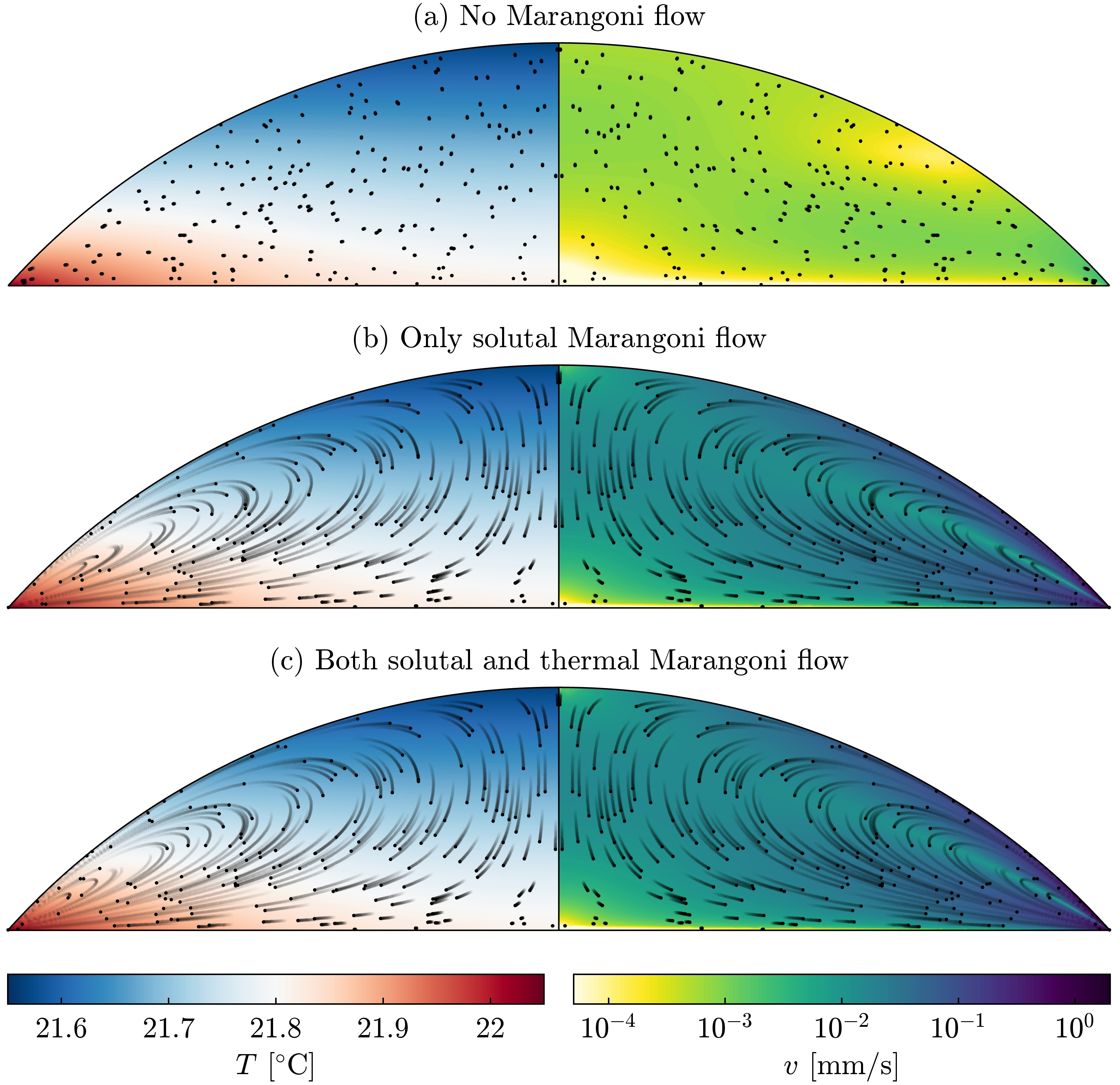}%
    \caption{Velocity and temperature in the drop at $t$ = 15 s with (a) no Marangoni flow, (b) only solutal Marangoni flow, and (c) both solutal and thermal Marangoni flow. The left half shows the temperature in the drop. The right half shows the velocity magnitude on a logarithmic scale. The trail of the tracer particle is 3 seconds long. The radius and height of the drop are $R = 0.576\ \mathrm{mm}$, and $H = 0.254\ \mathrm{mm}$}
    \label{fig:num_velocities}
\end{figure}

\begin{figure}
    \centering
    \includegraphics[width=0.80\textwidth]{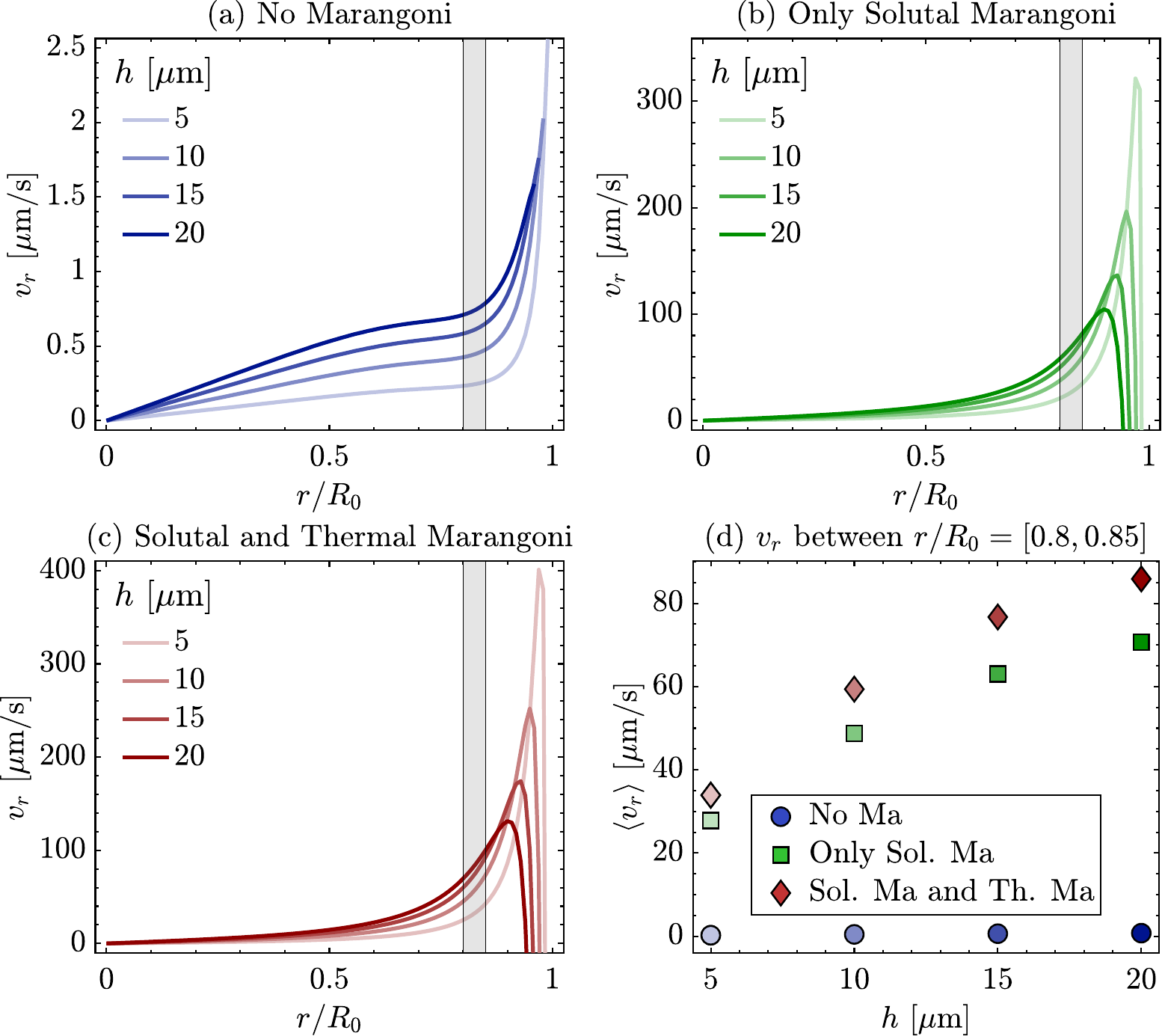}%
    \caption{(a,b,c) The radial velocity in the drop at different heights of the numerical simulations shown in \ref{fig:num_velocities}. The grey area is the interval over which the experiments have been averaged ($0.8<r/R_0<0.85$). (a) no Marangoni flow, (b) only solutal Marangoni flow, and (c) both solutal and thermal Marangoni flow. (d) The average numerical radial velocity averaged over the same interval as the experiments for the different drop heights. Note that the average radial velocity measured experimentally in \ref{fig:ptv} for an isolated drop is $v_0 = 15.8 \ \mathrm{\tcmu m/s}$.}
    \label{fig:num_ptv}
\end{figure}

To better understand the role of surface tension gradients due to the local composition and temperature we start with a constant surface tension before adding solutal and thermal effects. Figure \ref{fig:num_velocities}(a) shows the velocity and temperature in the drop without any Marangoni flow. Since the drop is pinned we observe a replenishing flow toward the contact line (i.e. coffee-stain flow), although the particle trails are not sufficiently long to adequately show the direction of the flow. The average velocity magnitude is 0.514 $\mathrm{\tcmu m/s}$. The temperature in the drop is overall 2.26 $\mathrm{^\circ C}$ colder than the ambient temperature due to evaporative cooling at the liquid-gas interface. Although the evaporative flux (and therefore the cooling) is higher near the contact line, this is actually the hottest part of the drop with the drop apex being the coldest part. This is due to the conductive substrate which supplies heat to the drop. In general, it is known that both the thermal properties and dimension of the substrate and the liquid drop will determine the temperature distribution in the drop \citep{ristenpart2007,dunn2009,diddens2017,wang2024apl}.

Next, we will make the surface tension dependent on the local composition (but not yet on temperature) as shown in figure \ref{fig:num_velocities}(b). We find that the flow is recirculating through the drop and that the average velocity magnitude is increased significantly to 26.3 $\mathrm{\tcmu m/s}$. Despite the large change in velocity, the average temperature as well as the temperature distribution is virtually identical compared to the case without Marangoni flow. Calculating the Péclet number using the average velocity, the radius of the drop and the thermal diffusivity of water, we find that 
\begin{equation}
 Pe_T = \frac{L v}{\alpha} = 1.06 \cdot 10 ^ {-7}.
\end{equation} 
This means that the heat transfer in the drop is dominated by conduction and is unaffected by the flow in the drop. Additionally, the source of heat (conduction through the glass slide) and the sick of heat (evaporative cooling at the liquid-gas interface) are also unchanged. Therefore the temperature in the drop is the same with or without solutal Marangoni flow.

Finally, we also make the surface tension dependent on the local temperature as shown in \ref{fig:num_velocities}(c). We find that the temperature distribution is identical to the previous cases and that the flow in the drop looks almost the same as the case with only solutal Marangoni flow. With thermal Marangoni, the average velocity magnitude in the drop is increased to 29.8 $\mathrm{\tcmu m/s}$, a difference of less than 12$\%$. Additionally, the direction of the thermal Marangoni flow is in the same direction as the solutal Marangoni flow, since it increases the velocity, meaning that we would expect that thermal effects enhance the attractive effect the drops have due to Marangoni contraction. 

To compare the numerical results more quantitatively with the experimental results we will evaluate the radial velocity numerically at the same heights as in the experiment. The depth of field (DoF) of the confocal is approximately 18 $\mathrm{\tcmu m}$ when the top surface of the substrate is perfectly in focus (see supplementary information). Figure \ref{fig:num_ptv}(a,b,c) shows the radial velocity at different heights for all three cases shown in figure \ref{fig:num_velocities}.

Without Marangoni flow, as previously mentioned we see the outward replenishing flow which diverges close to the contact line (fig. \ref{fig:num_ptv}(a)). With only solutal Marangoni enabled (fig. \ref{fig:num_ptv}(b)), we see that the radial velocity has a maximum and becomes negative due to the recirculation (only the positive part is shown in the figure). With both solutal and thermal Marangoni flow enabled (fig. \ref{fig:num_ptv}(c)), we see no change except in the magnitude of the flow. The shaded region corresponds to the interval over which the experiments have been averaged ($0.8<r/R_0<0.85$) and the numerical averages for each height on the same interval are shown in figure \ref{fig:num_ptv}(d).

The experimental radial velocity for an isolated drop is $v_0 = 15.8 \ \mathrm{\tcmu m/s}$. Which is significantly higher than the numerical radial velocity without Marangoni flow (0.515 $\mathrm{\tcmu m/s}$). This indicates that the flow in the experiments is dominated by Marangoni flow. However, the experimental radial velocity is not as high as the numerical radial velocity with solutal Marangoni flow (52.6 $\mathrm{\tcmu m/s}$). A similar deviation between experiments and numerics was found for the evaporation of water and glycerol drops by \citet{raju2022}.

Based on the numerical simulations we conclude that thermal Marangoni plays a minor role compared to solutal Marangoni flow. Qualitatively thermal and solutal Marangoni flow have the same effect since the direction of the thermal Marangoni flow is the same as the solutal Marangoni flow.


\section{Summary, conclusions, and outlook} \label{sec:conclusion}

In summary, we have experimentally studied evaporating neighbouring binary droplets, employing direct optical observations, confocal microscopy, and PTV. Our model droplets consisted of a water--1,2-hexanediol binary mixture, which is a standard model system for inkjet printing of multicomponent droplets. Due to the preferential evaporation of the droplets at the side away from the other droplets, we (and all colleagues we had asked) expected the centers of the drops to move towards each other. However, in contrast, they move apart. We could explain this behaviour as a consequence of the particles or contaminants in the drop being deposited on the sides away from the other drop, leading to preferential pinning on those sides, with all consequences for the overall flow dynamics in the droplets. We have shown that the flows in the neighbouring droplets become asymmetric due to the symmetry breaking by the other droplet, as expected, but the radial velocity close to the rim of the drop is still in good agreement with equation \ref{eq:f}, i.e., the local evaporative flux at the rim for a single component drop. Finally, with the help of direct numerical simulations for an evaporating isolated, axisymmetric binary droplet, we have elucidated the relative contributions of the replenishing flow and the solutal and thermal Marangoni flow, with the latter one playing only a minor role.

Though the model system consisting of water--hexanediol binary droplets is common used in the research of inkjet printing, it by far does not feature all possible effects. Different multicomponent liquid compositions would lead to different behaviour and in case of soluble or insoluble surfactants the situation become even more complicated. So our work is only the beginning, but given how counter-intuitive the observed behaviour was, we thought it was worth reporting and explaining our results. The relevance of such evaporating multicomponent neighbouring droplets in technological applications is huge, way beyond inkjet printing, and mastering their deposition and how they dry is key for most of these applications.


\section*{Declaration of Interests}
The authors report no conflict of interest.

\section*{Acknowledgements}
This work was supported by an Industrial Partnership Programme, High Tech Systems and Materials (HTSM), of the Netherlands Organisation for Scientific Research (NWO); a funding for public-private partnerships (PPS) of the Netherlands Enterprise Agency (RVO) and the Ministry of Economic Affairs (EZ); Canon Production Printing Netherlands B.V.; and the University of Twente. The authors thank Christian Diddens for discussions and setting up the numerical simulations; Lijun Thayyil Raju for discussions and preparing the OTS coated substrates; and Alvaro Marin for insightful discussions.


\appendix

\section{Numerical simulations of a droplet subjected to pure thermal Marangoni flow}\label{app:num_th_only}

\begin{figure}
    \centering
    \includegraphics[width=0.80\textwidth]{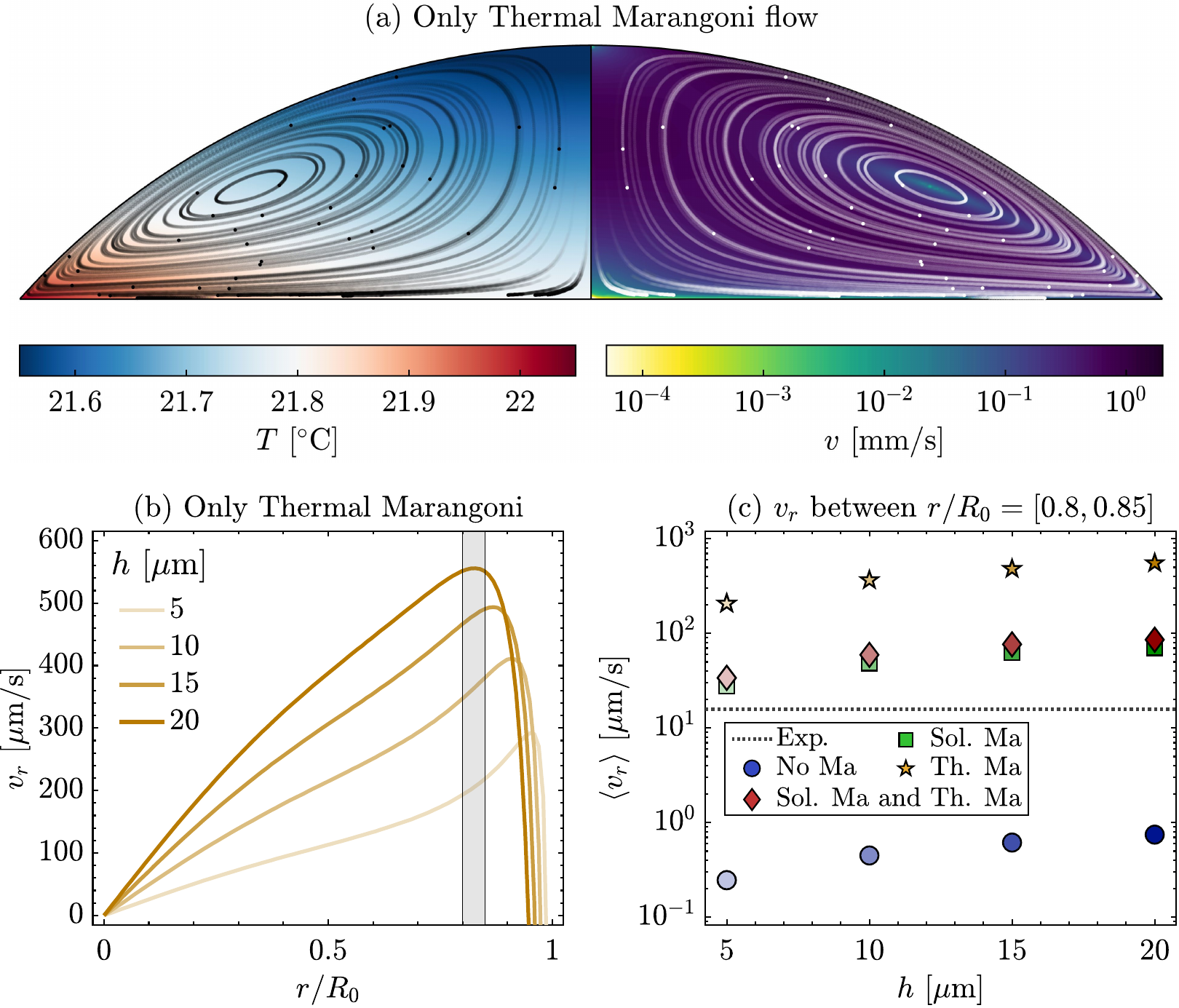}%
    \caption{(a) Velocity and temperature in the drop at $t$ = 15 s with only thermal Marangoni flow. The left half shows the temperature in the drop. The right half shows the velocity magnitude on a logarithmic scale. The trail of the tracer particle is 3 seconds long. For contrast, the colour of the trail is inverted on the right half of the drop. (b) The radial velocity in the drop at different heights of the numerical simulations shown in (a). The grey area is the interval over which the experiments have been averaged ($0.8<r/R_0<0.85$). (c) The average numerical radial velocity averaged over the same interval as the experiments for the different drop heights.}
    \label{fig:num_th_ma_only}
\end{figure}

\begin{figure}
    \centering
    \includegraphics[width=1\textwidth]{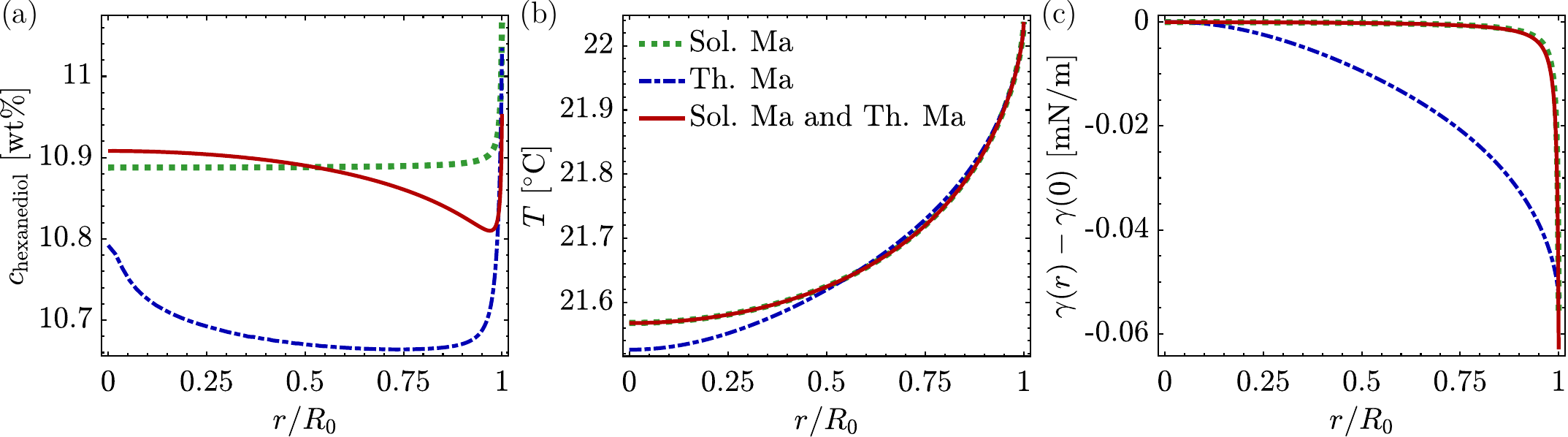}
    \caption{Various quantities evaluated at the liquid-air interface of the drop for only solutal, only thermal, and both solutal and thermal Marangoni flow. (a) The concentration of hexanediol (initially 10 $\mathrm{wt\%}$). (b) Temperature (ambient temperature is 24$\mathrm{^\circ C}$). (c) Surface tension $\gamma$ at $r$ minus the surface tension at the drop apex ($r=0$).}
    \label{fig:num_th_int}
\end{figure}

Figure \ref{fig:num_th_ma_only}(a) shows the numerical simulation where the surface tension only depends on temperature and not on the composition in the drop (i.e. only thermal Marangoni flow). Similar to the previous cases shown in figure \ref{fig:num_velocities}, the temperature distribution is identical. However, the flow in the drop is totally different. Not only are the trajectories of the particles very different, but the average velocity magnitude in the drop (593.3 $\mathrm{\tcmu m/s}$) is much higher that of only solutal Marangoni flow (26.3 $\mathrm{\tcmu m/s}$) and of solutal and thermal Marangoni flow (29.8 $\mathrm{\tcmu m/s}$). Figure \ref{fig:num_th_ma_only}(b) shows the radial velocity at different heights in the drop which are comparable to the DoF in the experiments. A comparison of all numerical velocities is shown in Figure \ref{fig:num_th_ma_only}(c) on a logarithmic scale. The numerical average radial velocity for the case with only thermal Marangoni flow is an order of magnitude larger than the case with both thermal and solutal Marangoni flow.

Since only solutal Marangoni flow and only thermal Marangoni flow are in the same direction, one might expect that solutal and thermal combined will enhance each other. Instead, we find enabling the solutal Marangoni effect almost nullifies the effect of thermal Marangoni flow. To unravel what causes this to happen we inspect the composition, the temperature, and the surface tension at the liquid-gas interface of the drop, as shown in figure \ref{fig:num_th_int}.

Examining the temperature at the drop surface (fig. \ref{fig:num_th_int}b), we see that the temperature gradient spans the whole drop. While the apex for the case with only thermal Marangoni flow is slightly colder, the magnitude of the temperature gradient is similar to the cases with solutal Marangoni flow. Therefore we expect that the change in surface tension due to the temperature is roughly the same in all cases. When we inspect the surface tension of only thermal Marangoni flow, we see that the surface tension changes across the entire drop. 

Now we turn to the case with only solutal Marangoni flow. We see that the concentration of hexanediol is roughly constant throughout the entire drop, except close to the contact line where it sharply increases due to selective evaporation. The concentration profile can be almost perfectly matched to the local evaporative flux of water which is also nearly constant throughout the drop but diverges near the contact line. The surface tension reflects the concentration and is constant except near the contact line. 

For the case with both solutal and thermal Marangoni flow we see, remarkably, that the surface tension gradient is almost identical to the of only solutal Marangoni flow. Looking at the composition we see that the concentration is no longer constant throughout the drop, but is slightly decreasing instead, before sharply increasing again at the contact line. This means that the solutal effect compensates perfectly for the thermal effect on the surface tension, except near the contact line.

This behaviour can be understood by considering that the solute (hexanediol) is transported by the flow in the drop, while the temperature remains unaltered. The thermal Marangoni flow transports the solute towards the apex of the drop where it accumulates (as can be seen in fig \ref{fig:num_th_int}(a) for only thermal Marangoni). When we enable solutal Marangoni flow, the surface tension gradients drives the solute away from the apex toward the water-rich area of the drop. We can confirm this by calculating the Péclet number for the transport of hexanediol in the drop using the average velocity (with both solutal and Marangoni flow), the radius of the drop, and the diffusion coefficient of hexanediol in water, we obtain
\begin{equation}
 Pe_c = \frac{L v}{D} = 0.011.
\end{equation} 
Although $Pe_c$ is still relatively small, it is five orders of magnitude larger than $Pe_T$ ($= 1.08\cdot10^{-7}$).

A similar observation has been made for the evaporation of sessile drop of pure water \citep{gaalen2022,rocha2024}, where a tiny amount of insoluble surfactant can completely suppress the thermal Marangoni flow. The difference between these past studies and the presented numerics here is twofold. One: in our case, the solute is soluble instead of insoluble. Two: for pure water drops it is nearly impossible to experimentally determine and quantify whether and how much surfactants are present on the surface of the drop. Whereas here we have a binary mixture where we know the initial composition precisely. However, that does not mean that we can rule out that some other type of insoluble surfactants or other contamination influences the flow in the drop. The deviation of the numerical from the experimental radial velocity for water/hexanediol and for water/glycerol \citet{raju2022} indicates that our understanding of these systems is not yet complete.

\newpage
\section{Supplementary Information}

\subsection{Setup}

\begin{figure}
    \centering
    \includegraphics[width=.8\linewidth]{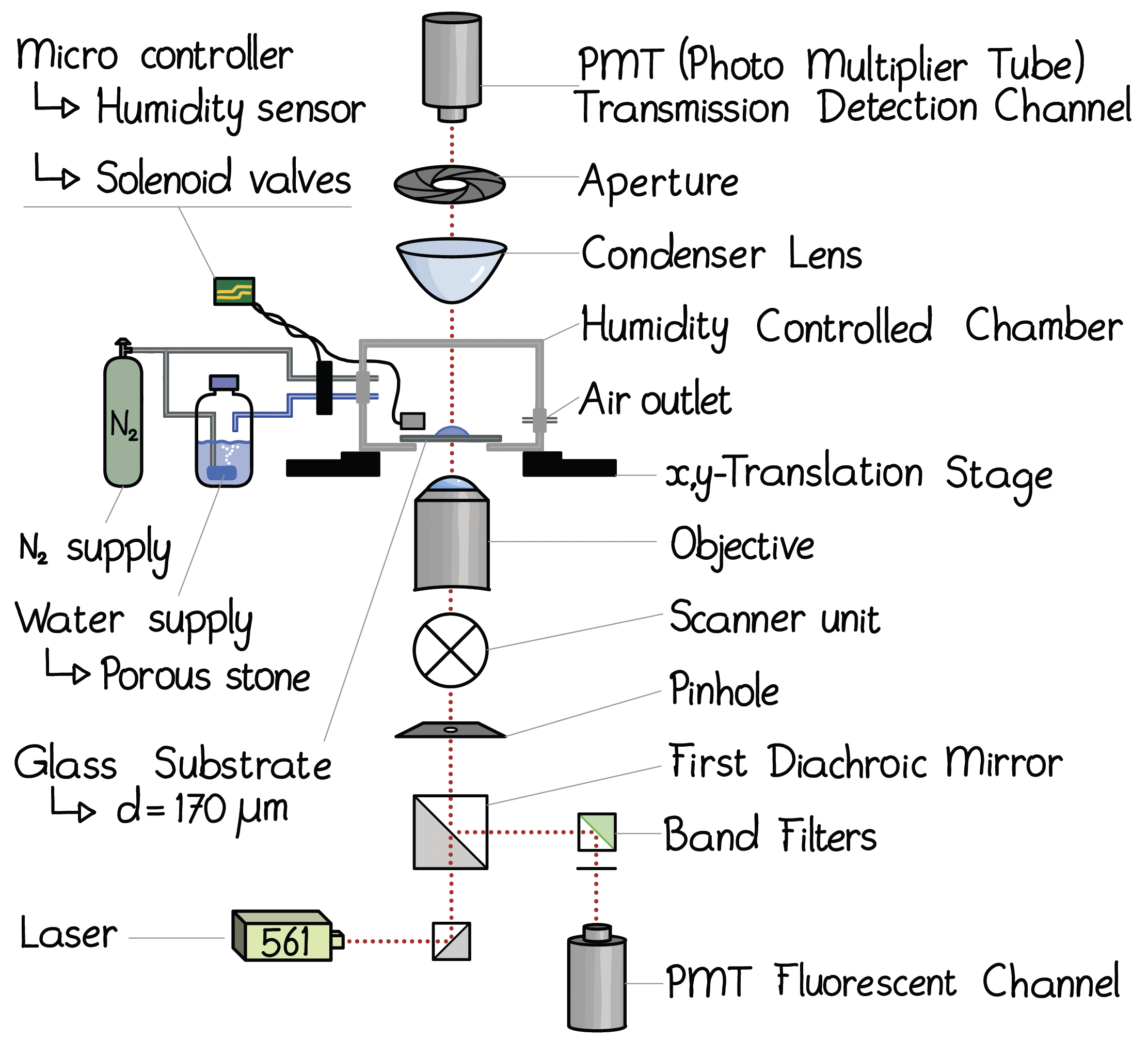}
    \caption{Schematic of the setup.}
    \label{fig:setup_confocal}
\end{figure}

\begin{figure}
    \centering
    \includegraphics[width=.8\linewidth]{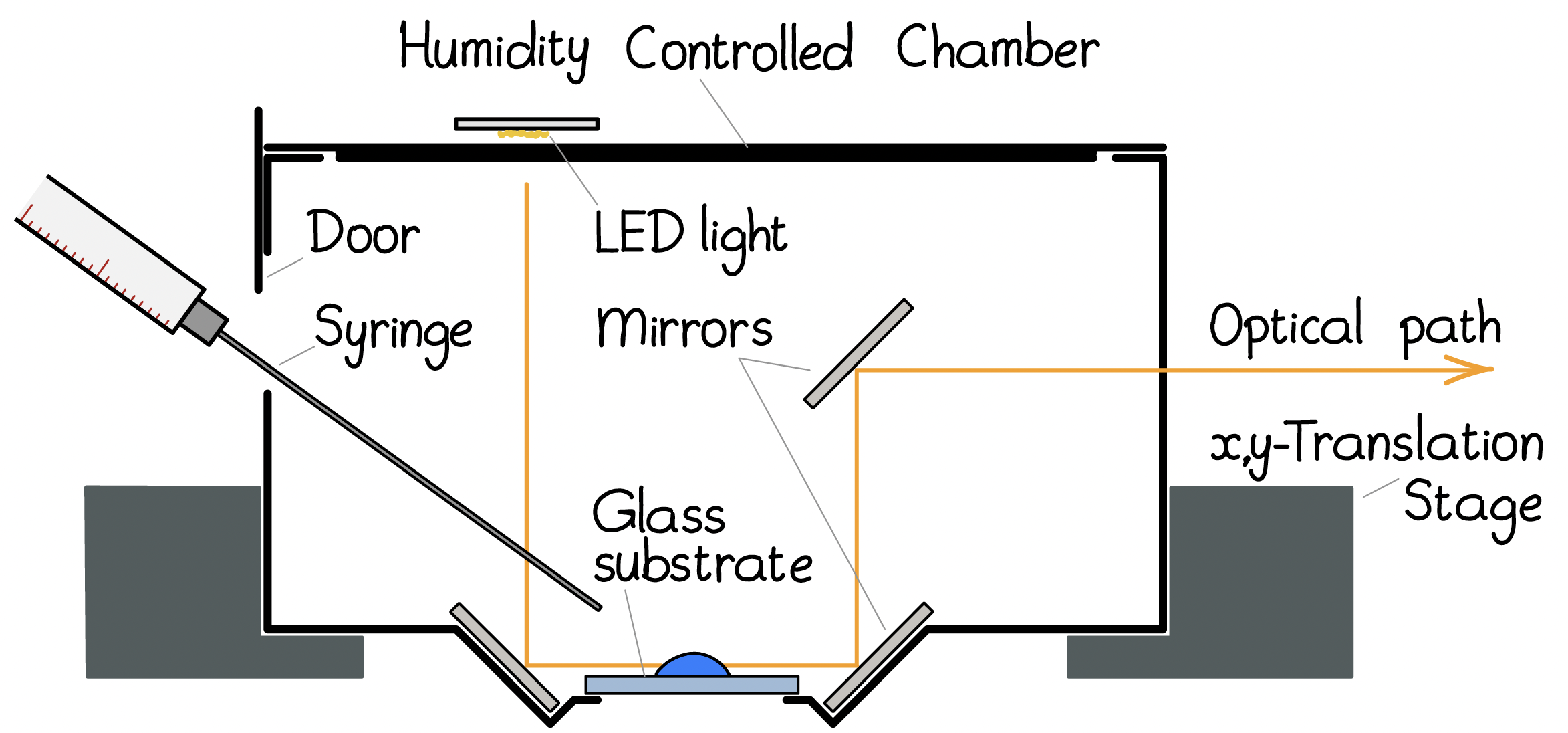}
    \caption{Schematic of the humidity chamber. (Not to scale)}
    \label{fig:setup_sideview}
\end{figure}

Fig. \ref{fig:setup_confocal} shows a schematic of the confocal microscope used for the measurements. To ensure that the experiments are repeatable and reproducible we had to monitor and control the ambient humidity and temperature in which the drop evaporates. To achieve this, we made a transparent acrylic chamber that fits inside the translation stage of the confocal. A schematic of the humidity chamber is shown in Fig. \ref{fig:setup_sideview}. To not interfere with the optics of the confocal microscope, there is no acrylic directly below the glass substrate. Unfortunately, this means that the chamber is not perfectly air-tight, and that the humidity will very slowly drift towards the ambient humidity in the lab. Additionally, at the start of each experiment the drops are deposited through a small door temporarily allowing some air in and out the chamber. Therefore, we have strived to keep the humidity in the chamber close to the humidity in the lab when doing experiments. 

The humidity is controlled using an in-house humidity controller, for which the schematics and codes are publicly available on GitHub\footnote{\url{https://github.com/Dennis-van-Gils/project-Humidistat}}. We monitor the humidity in the chamber with a digital sensor (Bosch Sensortec BME280) that measures humidity, temperature, and pressure. There are two sensors in the humidity chamber, one is very close to the substrate with the drop (1-2 cm) whereas the other sensor is as far away as possible from the drop (7-8 cm) in the upper corner of the humidity chamber. This allows us to assess the absolute and relative accuracy of the sensors, as well as the homogeneity of the relative humidity in the chamber. The sensors are connected to a microcontroller (Adafruit Feather M4 Express) which controls two solenoid valves. One is connected directly to the compressed nitrogen that is available in the lab (resulting in very dry air), the other passes first through a porous stone submerged in water (resulting in very humid air). Using this setup, we can achieve relative humidities between 15$\%$ and 85$\%$. During every experiment, both solenoid valves were closed to prevent any airflow in the chamber. 

The side view of the drop poses an extra challenge due to the requirements of the confocal and the humidity chamber. Namely, the substrate must be sufficiently close to the microscope objective to obtain an image. That means it is not possible to image the drop from the side directly because the view is obstructed by the translation stage (see Fig. Fig. \ref{fig:setup_sideview}). To solve this problem, we placed mirrors inside the chamber. However, this significantly increased the optical path length from the lens to the drop. This required a long-distance microscope lens with a sufficiently long working distance to be able to see the drop. The lens that we used is a modular Navitar lens with the following components: 2$\times$ F-Mount (1-62922), 12$\times$ Zoom (1-50486), 0.5$\times$ Lens Attachment (1-50012), Right Angle Lens Attachment (1-51490). This lens has a sufficiently large working distance of 165 mm, a maximum magnification of 7$\times$ and a resolve limit of 6.66 $\tcmu$m.

\subsection{Substrate preparation and characterisation: OTS coating}

The substrate was made hydrophobic by applying an hydrophobic coating using OTS \\(octadecyltrichlorosilane, $\mathrm{CH_3 (CH_2 )_{17} Si Cl_3}$). The mechanism and procedure are well described in \citet{silberzan1991}. OTS can covalently bond to the glass substrate using a silanisation reaction and will self assemble into a monolayer of alkanes, making the substrate hydrophobic. Note that water plays a crucial role in this reaction, as it is required in the silanisation reaction which bonds the OTS to the glass. However, when too much water is present, the OTS will start the reaction prematurely and react with itself. Therefore it is crucial prevent any presence of water. However, when too little water is present the reaction will not occur. This can be remedied by introducing a very small amount of water to the OTS solution by (1): letting the solution be exposed to humid air. Or using a more controlled method, (2): Adding water saturated chloroform/toluene solution. In our procedure the ambient humidity was sufficient to start the reaction.

We slightly deviated from the procedure outlined by \citet{silberzan1991}. Here are all the steps in detail:
\begin{enumerate}
    \item Prepare the substrate: Put the substrates in a glass substrate holder and fill with acetone. Sonicate (VWR, USC-T) the substrates for 5 minutes. Rinse with acetone, ethanol and water. Sonicate in Milli-Q water for 5 minutes. Dry with pressurised $N_2$. Treat substrate in plasma cleaner (Harrick Plasma, PDC 002) for 30 minutes at maximum power.

    \item Prepare the water-saturated chloroform/toluene solution: (In the fume hood!) Add 0.4 ml of water, 4 ml of chloroform, and 2 ml of toluene in a glass vail and close with a lid. Gently mix the solution by shaking. Let the vial rest for 2 h - 24 h, the water-saturated chloroform/toluene fill float on top of the remaining water.
    Using a pipet, carefully remove some (not all!) of the water-saturated solution without making contact with the water and put in another glass vial.

    \item Prepare OTS Solution: (In the fume hood!)
    \begin{itemize}
        \item \textit{With} the water-saturated chloroform/toluene solution: Add  48 ml of toluene, 12 ml of chloroform, and 1.2 ml of the water-saturated solution in a petri dish. Using a pipet, add 0.27 ml of OTS to the same petri dish. Immediately cover up with with aluminium foil to prevent water vapour from condensing into the solution and prematurely starting the reaction of OTS with itself.
        \item \textit{Without} the water-saturated chloroform/toluene solution: Add 25 ml of toluene in a petri dish. Using a pipet, add 0.1 ml of OTS to the same petri dish. Immediately cover up with with aluminium foil to prevent water vapour from condensing into the solution and prematurely starting the reaction of OTS with itself.
    \end{itemize}
    
    \item Put the substrates in the OTS Solution: (In the fume hood!) Bring the OTS solution to the plasma cleaner when this is finished and remove the foil. Using metal tweezers, take the substrates from the plasma cleaner and directly submerge it in the OTS solution. With multiple substrates, make sure that none of the substrates are overlapping. Cover up the petri dish with aluminium foil immediately and move it back to the fume hood. Let the substrate sit in the OTS solution for 20 minutes.

    \item Cleaning the substrates: (In the fume hood!) Fill a glass substrate holder with chloroform. Remove the foil from the petri dish and transfer the coated substrates to the filled substrate holder, while minimising the time the substrate is in contact with air. Sonicate for 4 minutes at half power and leave another 10 minutes in the chloroform. Transfer to another substrate holder filled with ethanol, if necessary, carefully remove any white hairs from the surface of the substrate with a cotton bud, then sonicate for 4 minutes. Take out the substrate and let them dry. Then rinse with ethanol and water and dry with compressed $N_2$.
\end{enumerate}

\begin{figure}
    \centering
    \includegraphics[width=.8\linewidth]{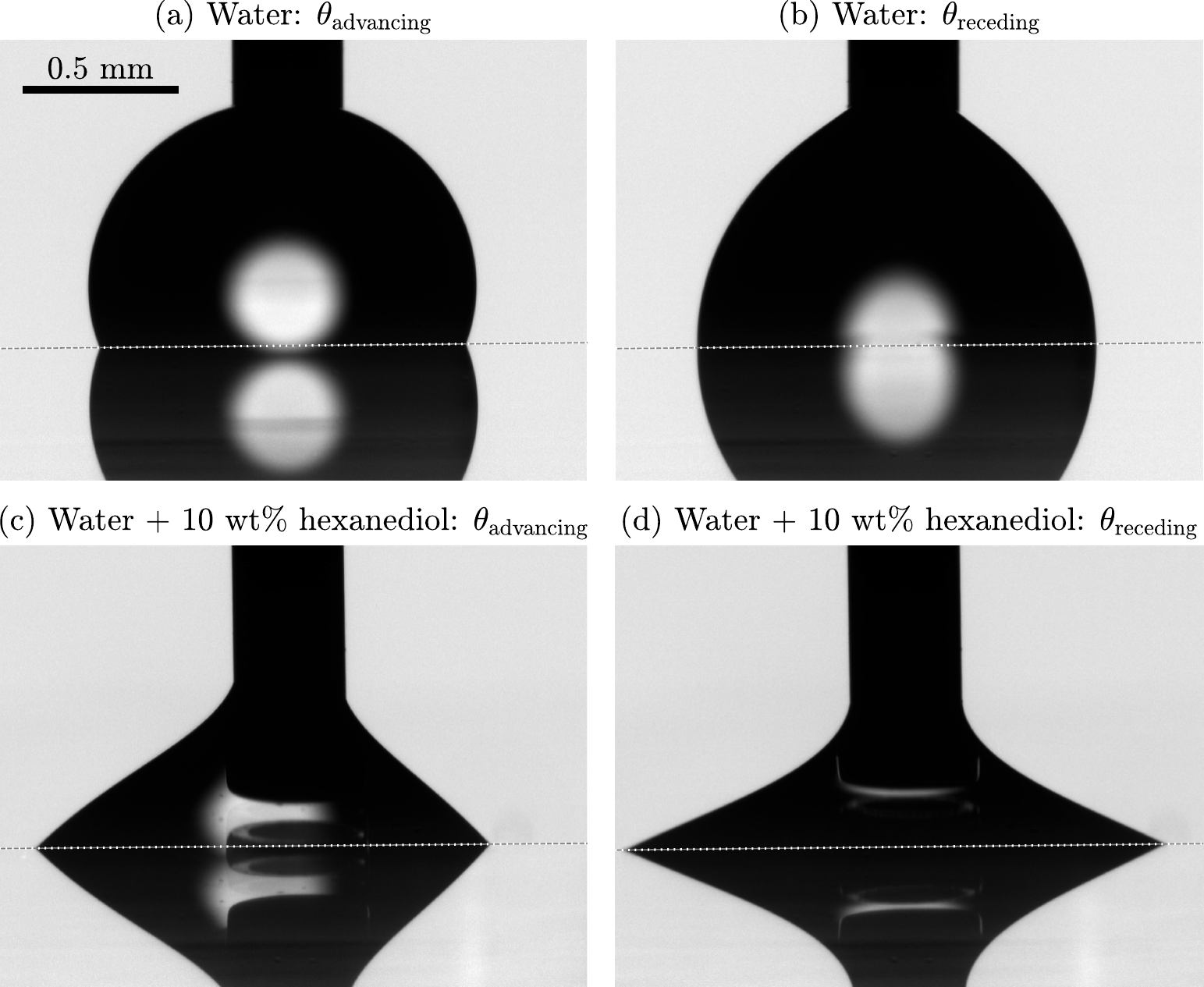}
    \caption{Contact angle of water (top row) and water + 10 wt$\%$ hexanediol (bottom row) on the OTS coated substrate. The left column shows the drops as they are slowly inflated and the right column as they are deflated. The scale identical for all images. The discontinuous line indicates the substrate position, the drop and needle visible below this line is a reflection.}
    \label{fig:contact_angle}
\end{figure}

\subsubsection{Contact angle measurement}
We measured the advancing and receding contact angle of water on the OTS coated substrate by slowly (0.1 $\tcmu$l/s) inflating and deflating a drop on the substrate. Using a side view camera and diffused backlight we imaged the drop. The reflection makes it simple to measure the contact angle. Figure \ref{fig:contact_angle} shows the experimental images captured of water and a mixture of water with 10 wt$\%$ 1,2-hexanediol. The measured contact angles are as follows:
\begin{alignat*}{4}
    \text{Water:}\ &\theta_\mathrm{advancing}\ &=\ &&110.3^\circ \pm 0.6^\circ\\
    \text{Water:}\ &\theta_\mathrm{receding}\ &=\ &&89.7^\circ \pm 1.1^\circ\\
    \text{Water\ +\ 10\ wt\%\ hexanediol:}\ &\theta_\mathrm{advancing}\ &=\ &&46.2^\circ \pm 0.7^\circ\\
    \text{Water\ +\ 10\ wt\%\ hexanediol:}\ &\theta_\mathrm{receding}\ &=\ &&24.8^\circ \pm 0.8^\circ
\end{alignat*}
We see that both water and water/hexanediol show contact angle hysteresis. Indicating that the the surface is not perfectly homogeneous or unavoidable contaminants are present. Note that the contact angle measurement was not performed in a humidity controlled environment and that therefore evaporation will inevitably impact the resulting measured contact angle to some degree, in particular for water/hexanediol. Nevertheless, this measurement does provide insight in the surface properties of OTS.


\subsection{Side view measurements}

\begin{figure}
    \centering
    \includegraphics[width=.7\linewidth]{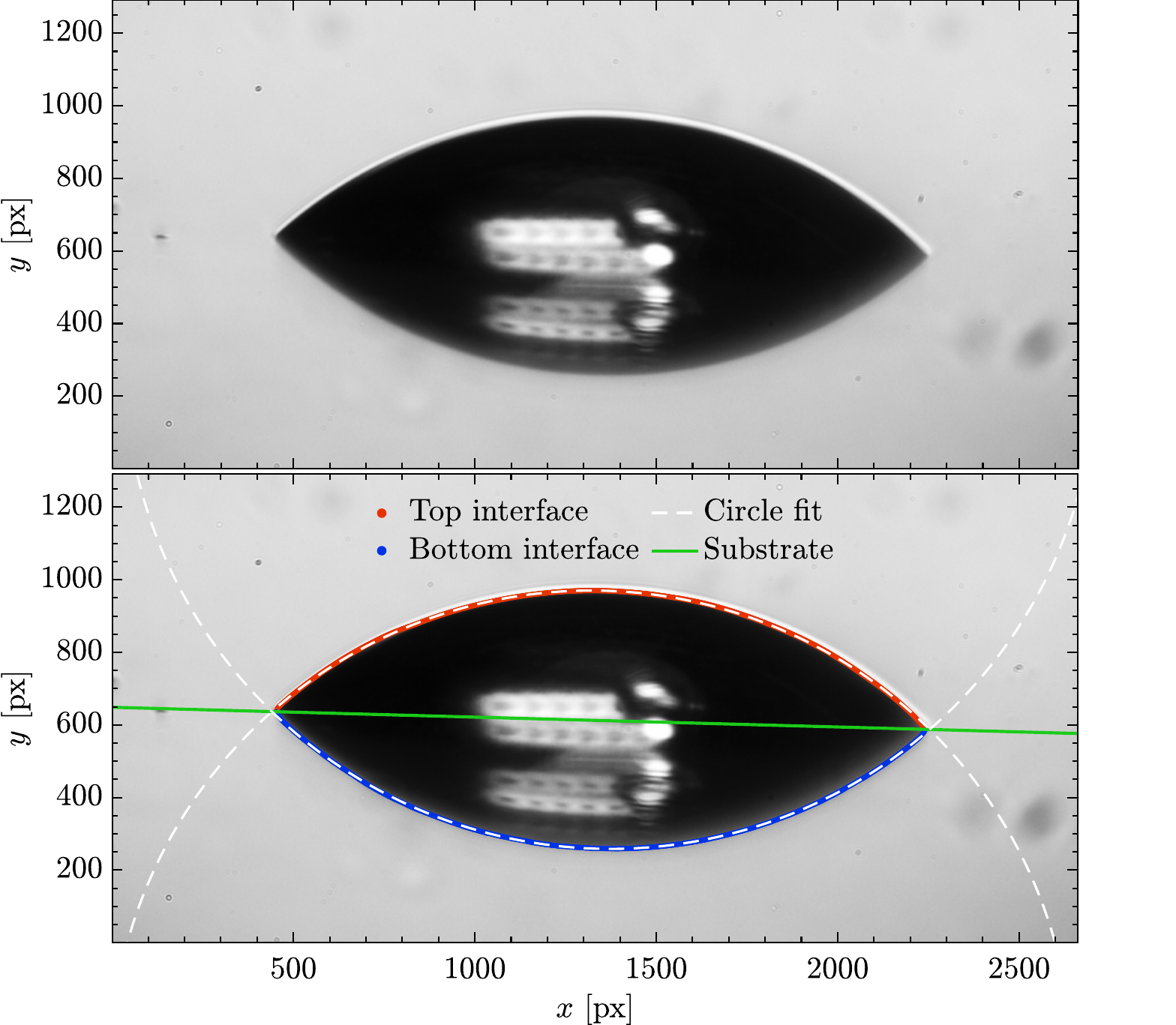}
    \caption{Top panel: Image captured by the side view camera. Bottom panel: Detected interface and the fitted circle of the side view image.}
    \label{fig:drop_image}
\end{figure}

The side view allows us to track the positions of the drop and to measure the drop volume over time. Fig. \ref{fig:drop_image}(a) shows a typical image captured by the side view camera.  Although the drop appears tilted, the substrate is perfectly level. The tilt is introduced due to the imperfect alignment of the mirrors. The side view is pitched 4$^\circ$ forward such that the reflection of the drop is visible in the substrate.

\subsubsection{Image processing}

Before processing the image, the intensities of the captured image are normalized between 0 and 1. Then we apply a Gaussian blur with standard deviation $\sigma = 2$ px. This reduces the noise in the image significantly without compromising the information of the drop interface, since the features of the drop are much larger then the applied blur. The system is lens limited since the resolution of the camera is 0.64 $\tcmu$m/px, which makes one pixel approximately 10 times smaller than the optical resolution of the lens. We distinguish the drop from the background by binarising the image using a threshold of 0.5. Using morphological operations we fill any speckles/reflections in the drop interior and the background. To identify the edge, we calculate intersection between the threshold and the intensities to obtain the interface with sub-pixel accuracy. We do this for  every column of pixels in the image, such that we ca easily differentiate between the top and bottom interface.

\begin{figure}
    \centering
    \includegraphics[width=.8\linewidth]{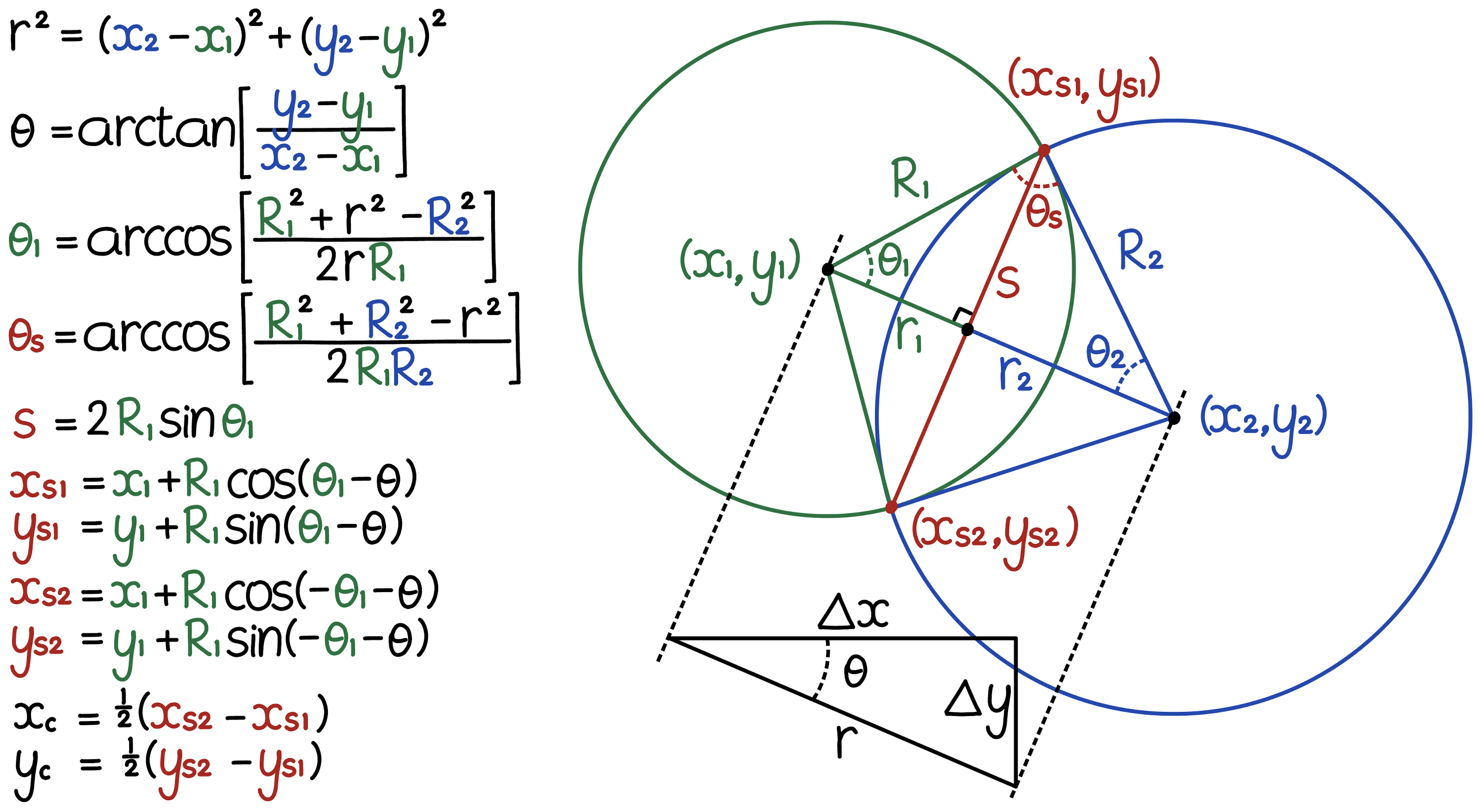}
    \caption{Geometry of two intersecting circles.}
    \label{fig:geometry}
\end{figure}

Next, we fit a circle to the top and bottom interface using the least squares method. This fit returns the radii ($r_1$, $r_2$) and the centers of the top and bottom circle respectively ($x_1$, $y_1$ and $x_2$, $x_2$). Using geometry (Fig. \ref{fig:geometry}) we can calculate the tilt of the camera, the contact angle of the drop, and the contact radius of the drop. The volume of a spherical cap is given by the following:
\begin{align}
    V = \frac{\pi R^3}{6}(2 + \cos{\theta})\frac{\sin{(\theta/2)}}{\cos^3{(\theta/2)}}
\end{align}
The results of the volume, contact radius, and contact angle for an isolated are shown in Fig. \ref{fig:drop_isolated}. Additionally the relative humidity and the temperature measured by the near- and far-away sensor are shown. As the drop is deposited, the sensors show a sharp increase in the humidity. The sensor close to the drop increases much more sharply than the sensor far from the drop. When the drop has evaporated, the humidity measured by the sensor close to the drop decreases, but the sensor far from the drop remains unaffected. This indicates that the chamber sufficiently large to be considered an infinite medium. 

\subsubsection{Diffusion limited evaporation model}

\begin{figure}
    \centering
    \includegraphics[width=1\linewidth]{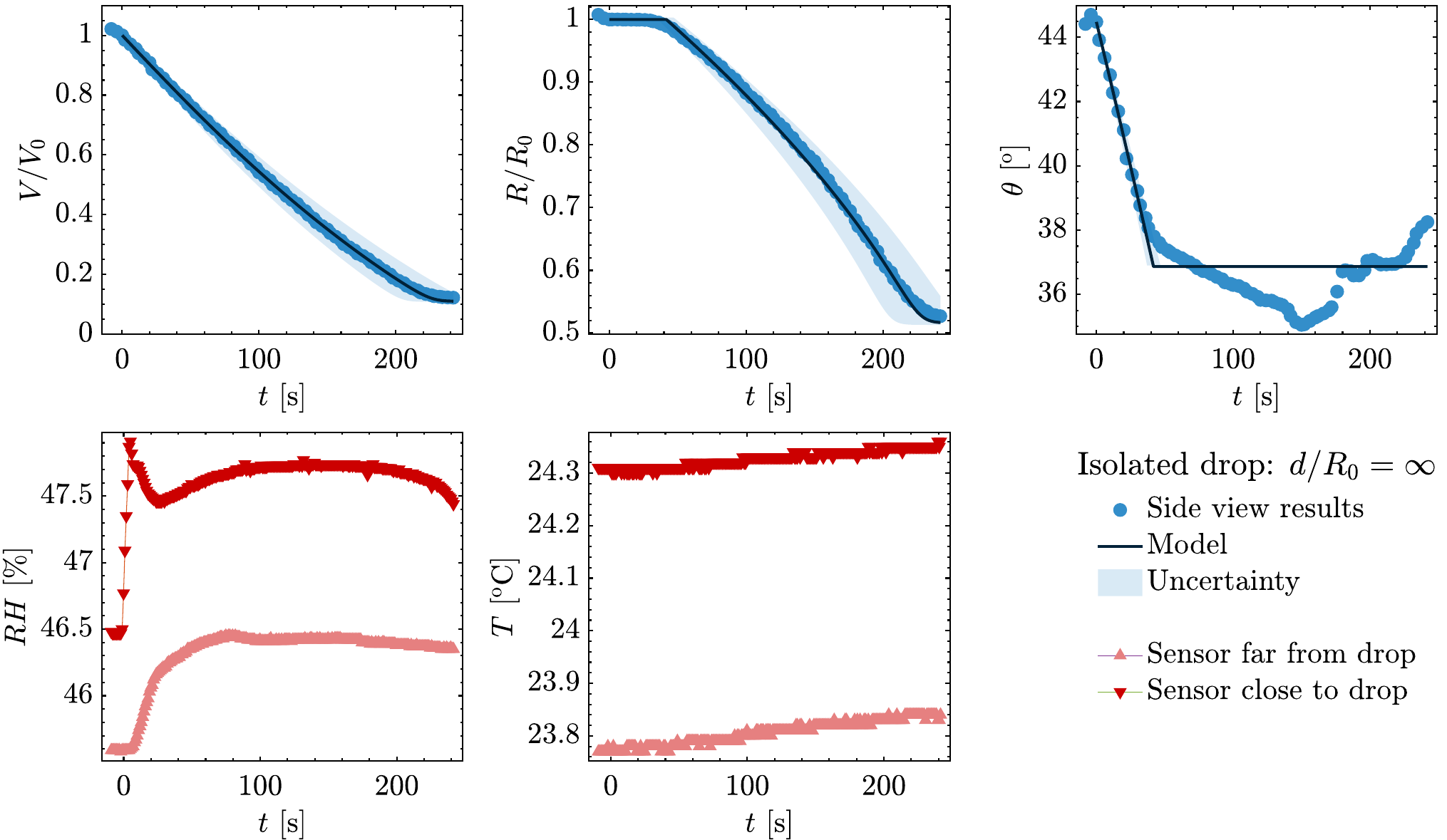}
    \caption{Isolated drop with initial 1,2-hexanediol concentration of 10 wt$\%$. Top row: evolution of the drop volume, contact radius, and contact angle during evaporation. The solid lines are calculated using the diffusion limited evaporation model where the initial conditions are based on the measurements. The shaded region is the variation in the model due to the uncertainty in the measured quantities. Bottom row: the measured humidity and temperature for the nearby sensor and the far-away sensor. Initial conditions: $V_0 = 0.135 \,\pm\, 0.002 \ \mathrm{\tcmu l}, \ R_0 = 0.583 \,\pm\, 0.002 \ \mathrm{mm}$.}
    \label{fig:drop_isolated}
\end{figure}

\begin{figure}
    \centering
    \includegraphics[width=1\linewidth]{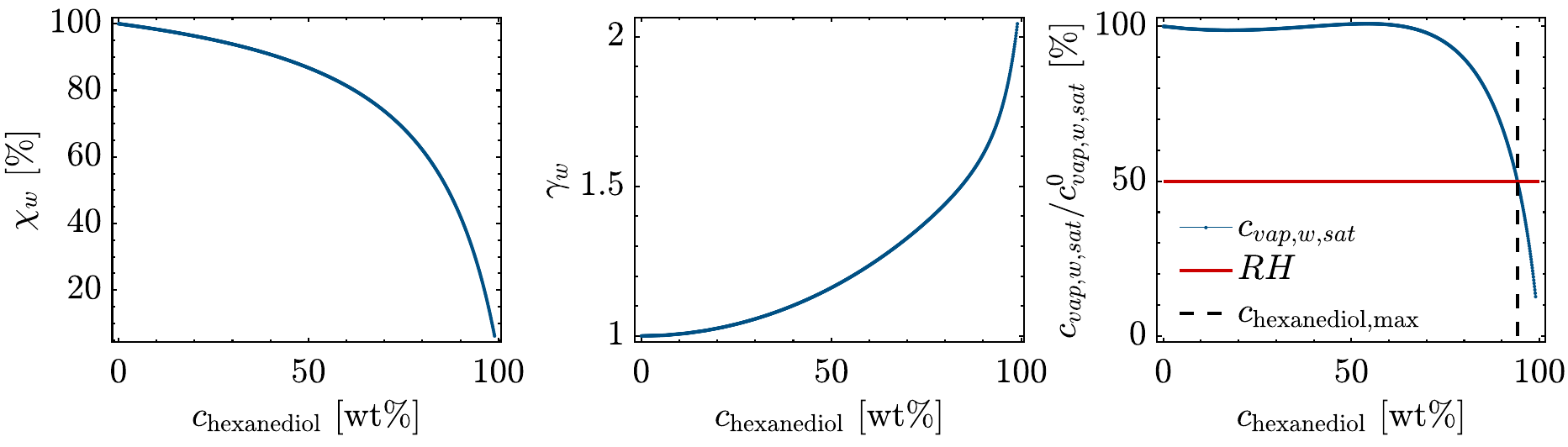}
    \caption{Left panel: \textit{mole} fraction of water versus the \textit{mass} fraction of hexanediol. Center panel: Activity coefficient of water versus the mass fraction of hexanediol. Right panel: normalized saturated vapour concentration of water versus the mass fraction of hexanediol. Up to $c_\mathrm{hexanediol}$ = 70 wt$\%$ the evaporation speed will remain close to that of pure water since $c_{vap,w,sat}/c_{vap,w,sat}^0 > 98 \%$. For a relative humidity of 50$\%$ the evaporation will cease at 94 wt$\%$ hexanediol, as is indicated by the dashed line.}
    \label{fig:vol_min}
\end{figure}

To further validate that the evaporation is indeed diffusion limited, we have also calculated and compared the volume evolution with the diffusion limited evaporation model \citep{popov2005}. In this model it is assumed that the transport mechanism that governs the evaporation speed is the diffusion of water vapour through the air and the mass flux is given by
\begin{align}
    \frac{dm}{dt} &= - \pi R D_{vap,w} \Delta c_{vap,w} g(\theta).
\end{align}
Here, $D_{vap,w}$ is the diffusion coefficient of water vapour in air, $\Delta c_{vap,w}$ the water vapour concentration difference, and $g(\theta)$ is a geometric factor that depends only on the contact angle and is given by
\begin{align}
    g(\theta) &= \frac{\sin{\theta}}{1 + \cos{\theta}} + 4 \int_0^\infty{\frac{1+\cosh{2\theta\tau}}{\sinh{2\pi\tau}}\tanh{[(\pi- \theta)\tau]}}d\tau.
\end{align}
The evaporation speed is directly proportional to the concentration difference in water vapour at the drop surface and at infinity. For a pure drop this is given by
\begin{align}
    \Delta c_{vap,w}^0 &= c_{vap,w,sat}^0(1-RH).
\end{align}
Here, $c_{vap,w,sat}^0$ is the saturated vapour concentration of pure water, and $RH$ the relative humidity. However, for a binary mixture the saturated vapour concentration is depends on the concentration of the other species according to Raoult’s law. For a binary mixture the concentration difference is given by
\begin{align}
    \Delta c_{vap,w} &= c_{vap,w,sat}^0(\chi_w \gamma_w - RH).
\end{align}
Here, $\chi_w$ is the mole fraction of water in the drop at the surface, and $\gamma_w$ the activity coefficient which takes the non-ideal interactions into account. $\gamma_w$ can be estimated using models such as the AIOMFAC model \footnote{\url{https://aiomfac.lab.mcgill.ca/index.html}} \citep{zuend2008}. When $\chi_w \gamma_w = RH$ the evaporation will cease, meaning that for a non-zero relative humidity not all water will evaporate as is further illustrated in Fig. \ref{fig:vol_min}. Note that we only have to consider the evaporation of water, since hexanediol is non-volatile. Additionally, we have assumed that the drop is well mixed at all times. A similar adaptation of the diffusion limited model has been used before to successfully describe the evaporation of a binary drop of water and glycerol \citep{raju2022}.

We evaluated the physical properties of the drop $2.26 \, \mathrm{^\circ C}$ lower than the ambient temperature to take evaporative cooling into account (based on the numerics presented in the appendix of the main text). This change in temperature predominantly has an effect on the vapour pressure of water, which becomes lower, extending the lifetime of the drop.

Now we still need to identify the contact line behaviour of the drop to fully describe the evaporation. Initially, the drop is pinned and evaporates in constant contact radius mode (CCR). When the contact angle reaches 38$^\circ$ the drop continues in constant contact angle mode (CCA). However, the contact angle isn't perfectly constant. A possible explanation is that the changing composition due to evaporation leads to a different equilibrium contact angle. In our model we reflect the contact line behaviour by starting in CCR mode until a critical contact angle of $\theta_c=37^\circ$ after which the drop continues is CCA mode.

The solid line in Fig. \ref{fig:drop_isolated} is the result of the model with the same initial conditions as measured in the experiment. The shaded region corresponds to the variation in outcome of the model due to the uncertainty in the measured experimental values. The model agrees very well with the measured volume. After approximately 240 s the evaporation ceases and the measured volume is 12.2$\%$ of the initial volume. The volume predicted by the model is at 11.1$\%$ of the initial volume.

\subsubsection{Shielding effect: Suppressed evaporation rate}

\begin{figure}
    \centering
    \includegraphics[width=1\linewidth]{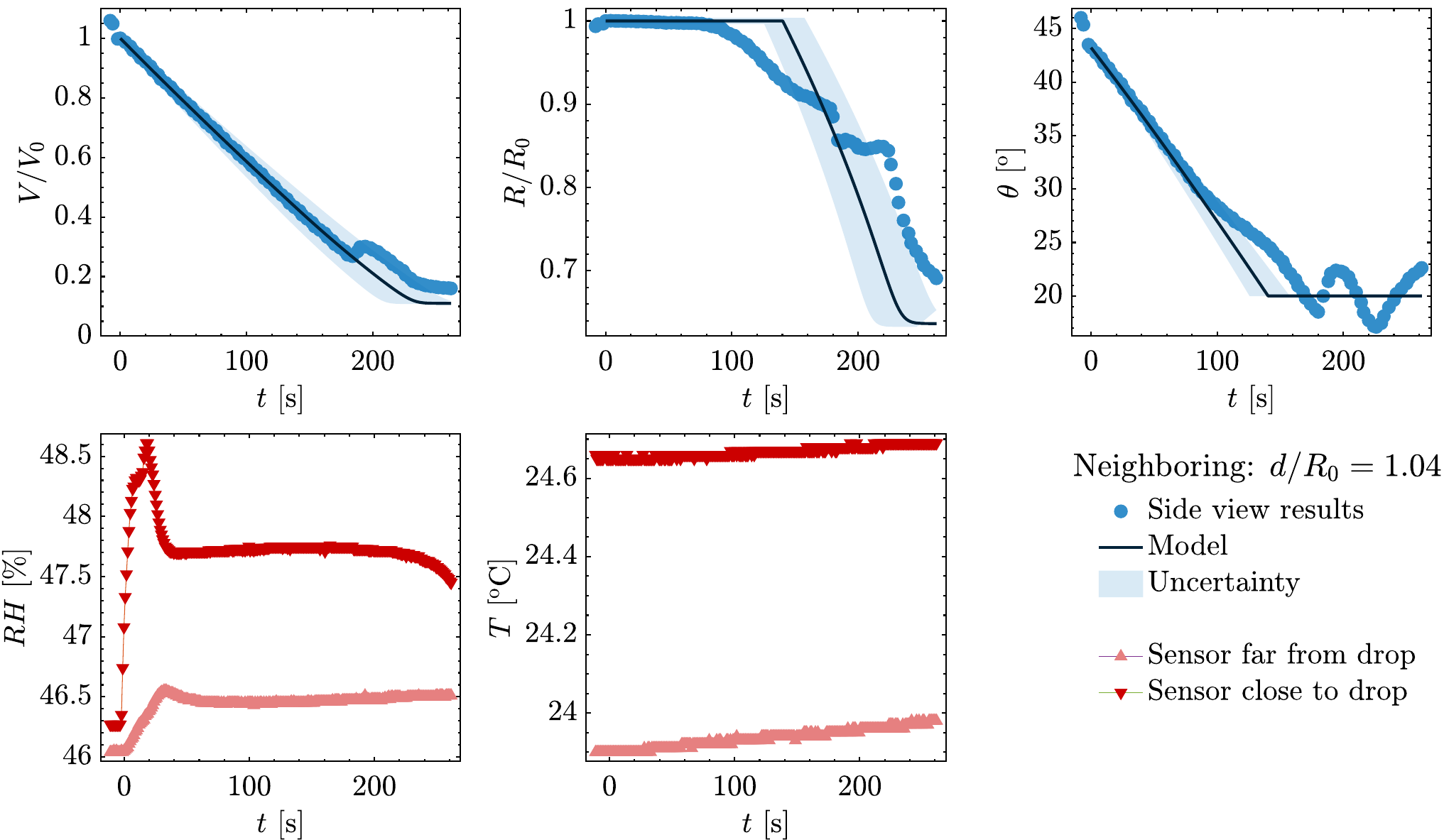}
    \caption{Neighbouring drop with initial 1,2-hexanediol concentration of 10 wt$\%$. Top row: evolution of the drop volume, contact radius, and contact angle during evaporation. The solid lines are calculated using the diffusion limited evaporation model where the initial conditions are based on the measurements. The shaded region is the variation in the model due to the uncertainty in the measured quantities. Bottom row: the measured humidity and temperature for the nearby sensor and the far-away sensor. Initial conditions: $V_0 = 0.124 \,\pm\, 0.002 \ \mathrm{\tcmu l}, \ R_0 = 0.573 \,\pm\, 0.003 \ \mathrm{mm}$.}
    \label{fig:drop_neighbouring}
\end{figure}

\begin{figure}
    \centering
    \includegraphics[width=.55\linewidth]{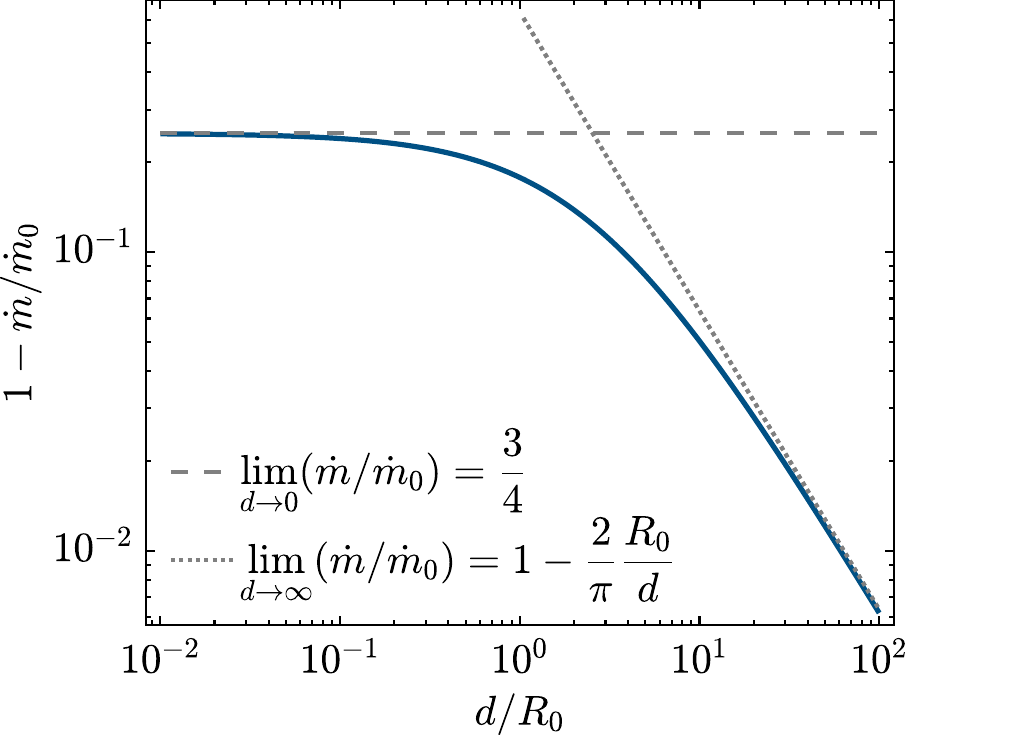}
    \caption{Reduction in the total mass flux due to the shielding effect of an identical neighbouring drop with initial contact radius $R_0$ at distance $d$. In the limit of touching drops ($d \ll R_0$), both drops evaporate 25$\%$ slower. When the distance between the drops is large ($d \gg R_0$), the shielding effect of the neighbouring drop diminishes as $d^{-1}$.}
    \label{fig:shielding_plot}
\end{figure}

Fig. \ref{fig:drop_neighbouring} shows the volume evolution for a neighbouring evaporating drop. In order to apply the diffusion limited model we need to consider the shielding effect of the neighbouring drop. For two identical pure drops with initial contact radius $R_0$ and separated by distance $d$ in the limit of $\theta \rightarrow 0$ and $d/R_0 \rightarrow \infty$ the integral shielding factor $F$ is given by \citep{fabrikant1985,wray2020}
\begin{align}
    F = \frac{\dot{m}}{\dot{m_0}} &= \displaystyle{\frac{1}{1 + \dfrac{2}{\pi}\arcsin{\left(\dfrac{R}{d+2R_0}\right)}}}.
\end{align}
Here, $\dot{m}_0$ is the mass flux for an isolated drop and $\dot{m}$ is the mass flux for a neighbouring drop. Fig. \ref{fig:shielding_plot} shows the shielding factor as a function of drop separation. Comparing the model with the experiment we find good agreement. Note that the apparent increase in volume at $t = 200$ s is due to the contact line behaviour where the drop shape temporarily deviates from a spherical cap. 

By applying Raoult's law and the shielding factor we can predict the volume evolution of evaporating neighbouring binary drop very well using the diffusion limited evaporation model.

\subsection{Particle Tracking Velocimetry (PTV)}

The defining feature of the confocal is its very narrow depth of field (DoF). By focusing at the substrate we can observe and follow small fluorescent particles that are carried with the flow, allowing us to visualize and quantify the flow. In this section we will describe the method and processing that we used to obtain quantitative information of the flow in the drop.

The confocal settings were as follows: Objective: magnification $ = 10 \times$; NA =  0.3; optical resolution: 1.03 $\tcmu$m. The scan area was 512 by 512 pixels, resulting in a pixel size of $2.50 \ \tcmu$m, and a field of view of 1.28 mm. The confocal scan mode was set to resonant mode. To increase the signal to noise ratio, we average every line 4 times, resulting in 7.7 frames per second. The pinhole size was: 29.4 $\tcmu$m. Laser wavelength = 561 nm, detection range: 570 nm – 620 nm. 

\begin{figure}
    \centering
    \includegraphics[width=.8\linewidth]{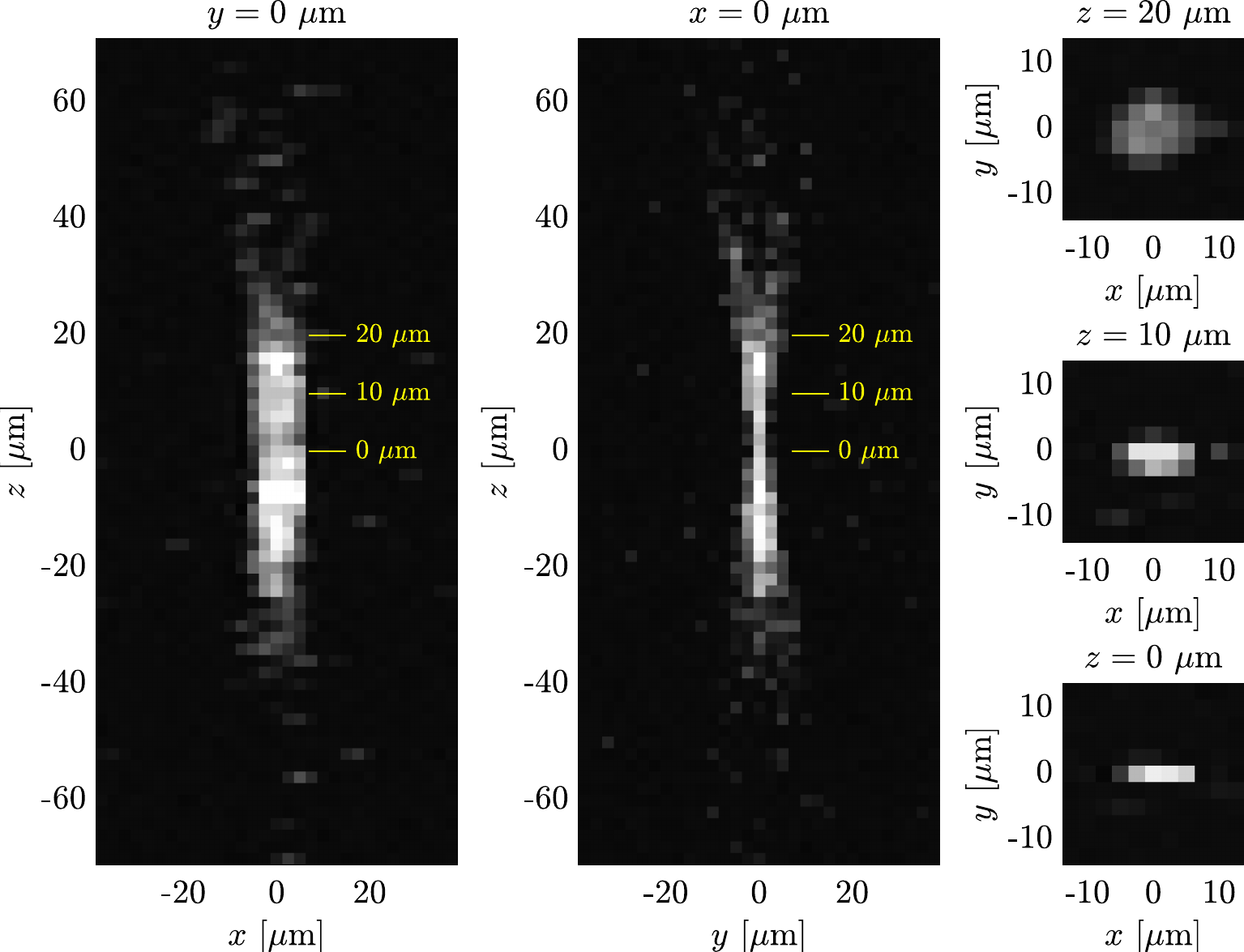}
    \caption{Depth of field of the confocal microscope at 10$\times$ magnification and $NA = 0.3$ and fluorescent particles with a diameter of $d=1.14 \ \tcmu$m. Slices in different directions are shown of a single particle.}
    \label{fig:dof}
\end{figure}

To estimate the effective DoF (depth of field) of the confocal microscope we scanned a single fluorescent particle at different heights. Fig. \ref{fig:dof} shows the slices of the particle in different planes. We can observe that the particle appears to be much larger in the $y$-plane as compared to the $x$-plane, this is an optical artifact of the scanning mode of the confocal. We see beyond $z = 20 \ \tcmu$m that the particle almost completely faded and we only measure some noise.

\subsubsection{Particle detection}

\begin{figure}
    \centering
    \includegraphics[width=1\linewidth]{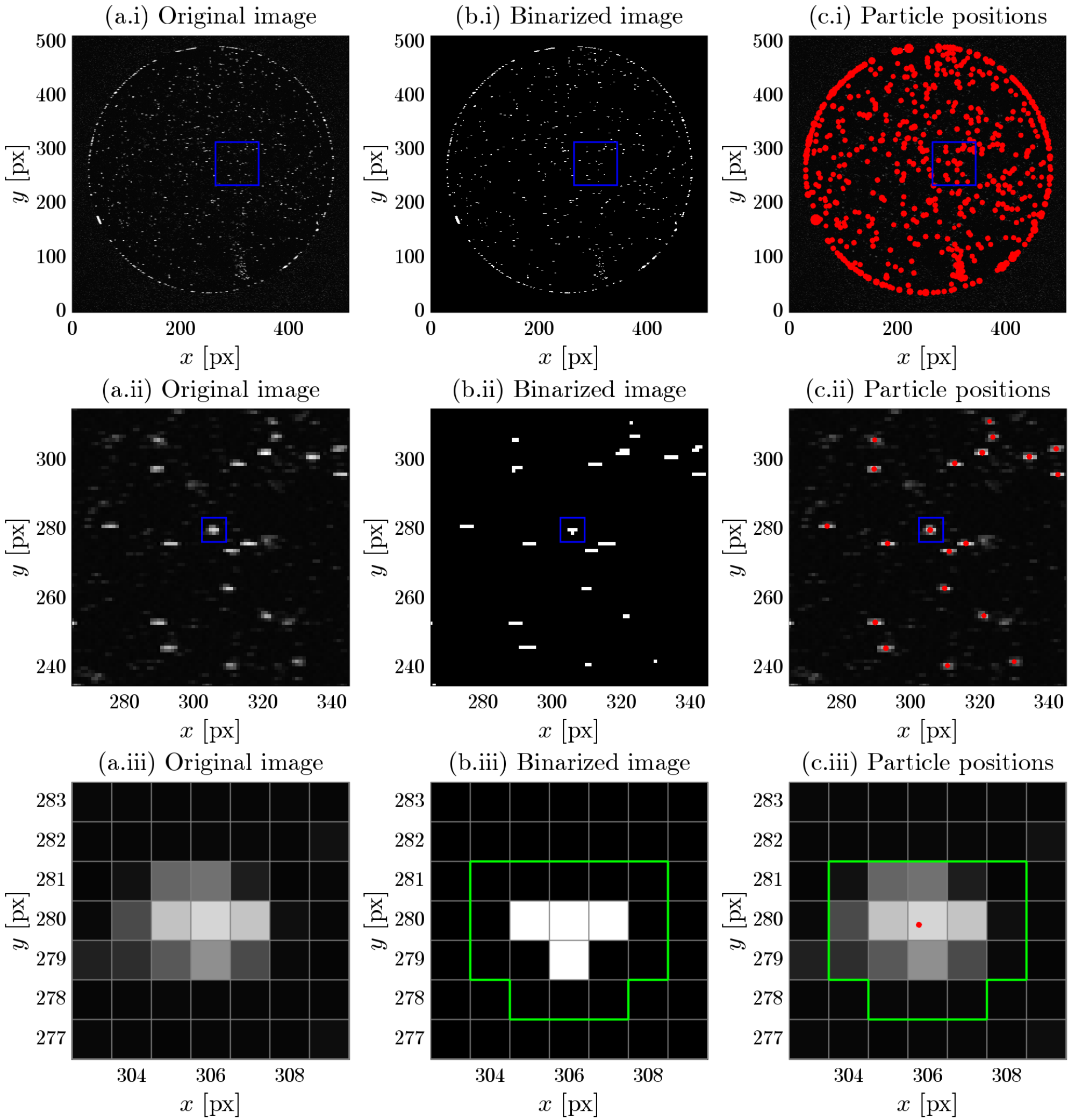}
    \caption{Particle detection. Columns: (a) original image as captured by the confocal; (b) image binarised using threshold; (c) original image with the detected particle positions. Rows: (i) full image; (ii) center of the image; (ii) single particle. The zoomed-in sections are highlighted with a blue box. The green line (b.iii) and (c.iii) indicates the region used to calculate the particle position and particle size.}
    \label{fig:ptv_detect}
\end{figure}

Fig. \ref{fig:ptv_detect}(a.i) shows an image captured by the confocal. Before processing, the intensities are normalized between 0 and 1. To detect the particle positions we apply a threshold of 0.45 resulting in a binary image as shown in Fig. \ref{fig:ptv_detect}(b). Each region of connected white pixels corresponds to a particle. To accurately determine the position of the particle we use the intensity of the original image, not the binarised image. First, we expand the region to include by 1 pixel, such that we do not miss neighbouring pixels that are just below the threshold but still relevant to the particle position. Fig. \ref{fig:ptv_detect}(b.iii) shows the expanded particle region with the green line.

Next, the particle position $(p_x, p_y)$ is determined by calculating the `center of intensity' region (analogous to the `center of mass') for each pixel. This is given by the following relations
\begin{align}
p_x &= \dfrac{\sum x_i I(x_i,y_i)}{\sum I(x_i,y_i)},\\
p_y &= \dfrac{\sum y_i I(x_i,y_i)}{\sum I(x_i,y_i)}.
\end{align}
Here, we sum over all the pixels in the particle region. The resulting particle positions are shown in Fig. \ref{fig:ptv_detect}(c). Note that this method can detect a displacement much smaller than a single pixel. Additionally, by using the intensities in the original image we can determine the particle positions much more accurately as compared to the binary image, which contains far less information.

\begin{figure}
    \centering
    \includegraphics[width=.8\linewidth]{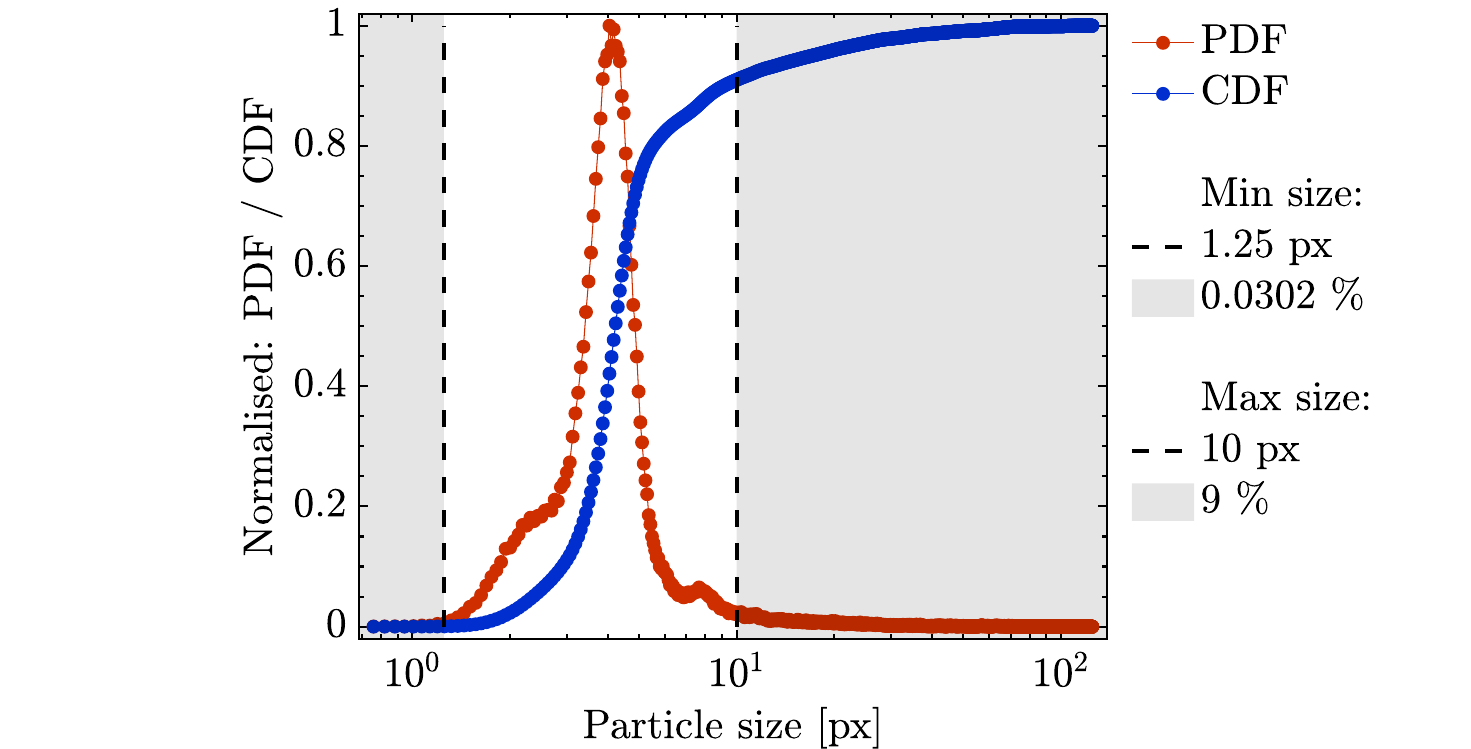}
    \caption{Probability density function (PDF) and cumulative probability density function (CDF) of the particle sizes (sum of the intensities). The regions highlighted in gray are limiting particle sizes below and above which are omitted. The percentage of eliminated particles is shown in the legend.}
    \label{fig:ptv_size}
\end{figure}

Finally, we filter out particles that are incorrectly detected as particles. We do this based on the size of the particles, which we calculate using,
\begin{align}
p_{size} =  \frac{\sum I(x_i,y_i)}{I_0}.
\end{align}
This is simply the sum of all the intensities of the pixels in the particle region. This is very similar to a `particle mass'. Note that this is \textit{not} a particle radius. When a particle size is very small, it is much more likely to be noise. When a particle size is very large, it is likely a cluster of multiple particles that are very close or even overlapping. Fig. \ref{fig:ptv_size} shows the probability density function of the particle size distribution. All particles smaller than 1.25 or larger than 10 are excluded.

\subsubsection{Particle matching}

Now that we have the particle positions for each frame, we need to match the detected particles between the different frames to reconstruct their trajectories. Since the displacement of the particles between consecutive frames is very little, we can use the nearest-neighbour approach. For each particle in frame $n$ we calculate the distance to all the other particles in frame $n+1$. We then match the particle with the smallest distance.

To prevent mismatches, we apply the following criteria.
\begin{itemize}
    \item  If the distance between the particles is 5 pixels or larger, it is very likely to match the wrong particle. 
    \item If multiple particles are found within a 1-pixel radius, we cannot distinguish the particles well enough to match them with certainty.
    \item  If a particle in frame $n+1$ has been matched to multiple particles in frame $n$, we also disregard those matches.
\end{itemize}
The small displacement of particles between frames makes the matching process computationally very straightforward. With the above criteria we loose some information when multiple particles are in close proximity, although more advanced methods the can distinguish overlapping particles, with the current method we retain a sufficient amount of information for analysis. For all successfully matched particle trajectories we require that the minimum trajectory length is at least 15 frames. Due to the large frame rate we need this many frames to obtain any relevant displacement to ensure accurate velocities measurements.

Due to noise in the image and naturally present Brownian motion of the particles, the velocities that we measure are very noisy. To eliminate some of the noise without compromising the signal we apply a  filter, which is a moving average with the following kernel: $[1/4, 1/2, 1/4]$. This filter is applied to both the positions of the particles before calculating the particle velocities, as well as after calculating the particle velocity. Note that the filtering we apply is minimal and that after filtering the data is still very noisy. This is a deliberate choice, since we will bin and average the data later. Keeping the noise allows us to quantify the noise/Brownian motion of the particles.

\subsubsection{Azimuthal dependence of the radial velocity}

\begin{figure}
    \centering
    \includegraphics[width=.9\linewidth]{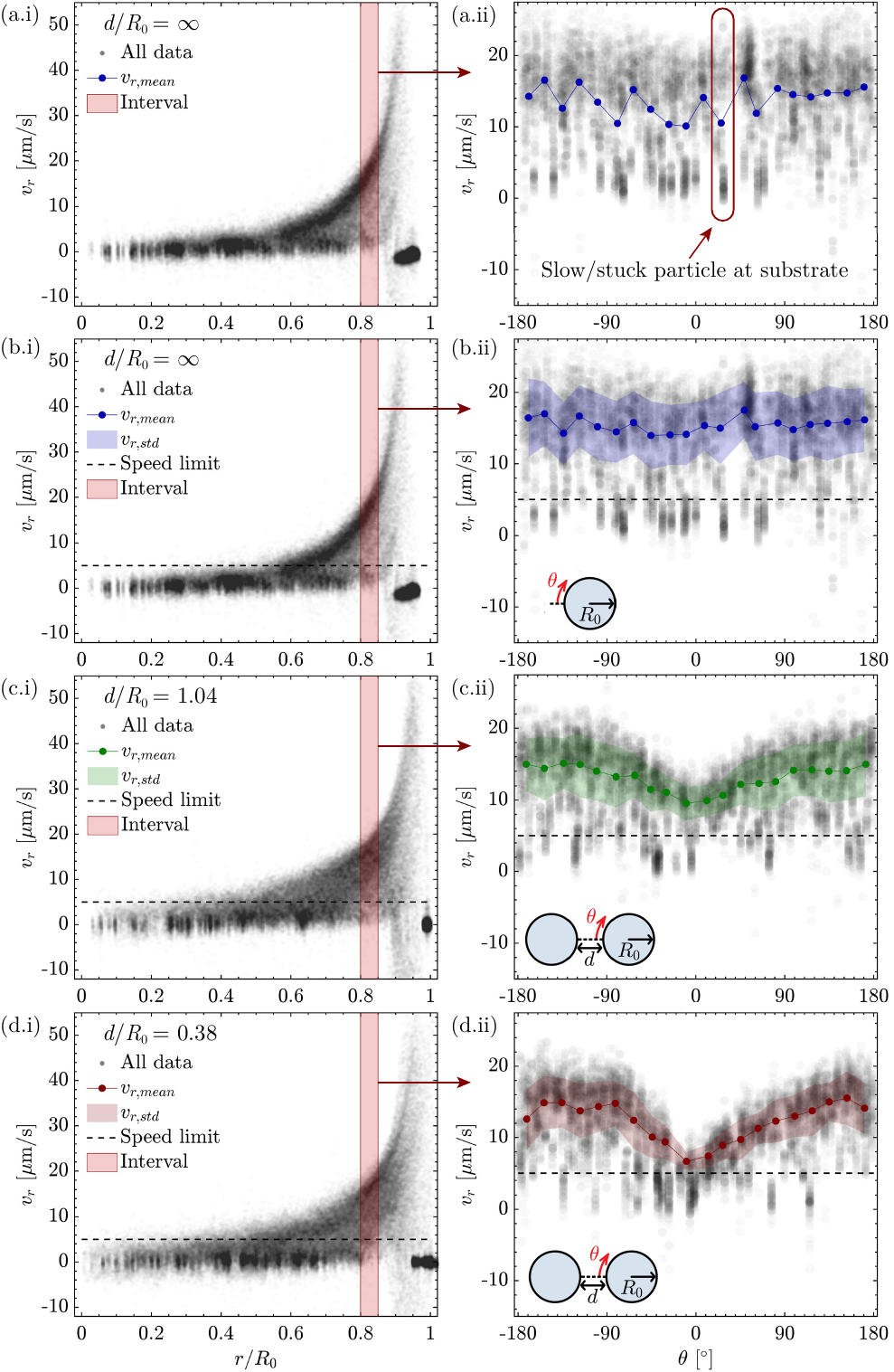}
    \caption{Radial velocity measurements using PTV. Column (i): All radial velocities $v_r$ versus the radial position $r/R_0$ in the drop. Column (ii): Radial velocities $v_r$ on the interval $0.8<r/R_0<0.85$ versus the azimuthal position $\theta$ in the drop. Row (a): Isolated drop \textit{without} velocity limit. Row (b): Isolated drop \textit{with} velocity limit. Row (c): neighbouring drop at distance $d/R_0 = 1.04$. Row (d): neighbouring drop at distance $d/R_0 = 0.38$}
    \label{fig:velo_lim_theta}
\end{figure}

Now that we have calculated all the trajectories and velocities of all the particles in the drop we can extract the azimuthal dependence of the radial velocity close to the contact line for all the measurements. Fig. \ref{fig:velo_lim_theta}(a.i) shows all measured radial velocities versus the radial position for an isolated drop. Close to $r/R_0 = 0$ the velocities are below $0.5 \ \tcmu$m/s which is likely Brownian motion. As we approach the contact line (i.e. $r/R_0 = 1$), the measured radial velocities start to increase. After $r/R_0 = 0.6$ we see two branches of radial velocities. One velocity-branch continues to increase approaching the contact line. However, the other velocity-branch remains close to $v_r = 0$. 

Next we take all the radial velocities that we measured between 0.8 and 0.85 and plot those velocities as a function of the azimuthal position in the drop as shown in Fig. \ref{fig:velo_lim_theta}(a.ii). Here we can observe again a split in the measured velocities, there are a large number of measurements around $v_r = 2 \ \tcmu$m/s and concentrated patches of measurements at $v_r = 0 \ \tcmu$m/s. Our interpretation is that some particles are strongly interacting with the substrate and are no longer acting as reliable tracer particles. Whereas other particles, which are slightly further away from the substrate, are still valid tracer particles. Unfortunately, the particles which are slow/stuck have a large impact on the average radial velocity since they are always between $0.8<rR_0<0.85$. In contrast, the particles following the flow are only briefly in the same window.

To be able to distinguish between the slow/stuck particles and the valid tracer particles, we discriminate between them using a “minimum speed limit". Which was manually determined by eliminating the individual particles as much as possible without interfering with the rest of the measured particles. Fig. \ref{fig:velo_lim_theta}(b.ii) shows the result when we apply a speed limit of $5 \ \mathrm{\tcmu m/s}$ on the azimuthal distribution of the radial velocity. The sharp spikes are now filtered out. Fig. \ref{fig:velo_lim_theta}(c) and (d) show the radial velocities for neighbouring drops. Both suffer from the same problem of slow/stuck particles near the substrate. Note that without applying the speed limit the data becomes a lot more noisier, but the results and conclusions in the main text remain the same. 

\newpage
\bibliographystyle{jfm}
\bibliography{main}

\begin{thebibliography}{52}
\expandafter\ifx\csname natexlab\endcsname\relax\def\natexlab#1{#1}\fi
\def\au#1{#1} \def\ed#1{#1} \def\yr#1{#1}\def\at#1{#1}\def\jt#1{\textit{#1}} \def\bt#1{#1}\def\bvol#1{\textbf{#1}} \def\vol#1{#1} \def\pg#1{#1} \def\publ#1{#1}\def\arxiv#1{#1}\def\org#1{#1}\def\st#1{\textit{#1}}

\bibitem[Brinker {\em et~al.\/}(1999)Brinker, Lu, Sellinger \& Fan]{brinker1999}
{\sc \au{Brinker, C.~J.}, \au{Lu, Y.}, \au{Sellinger, A.} \& \au{Fan, H.}} \yr{1999}  \at{Evaporation-induced self-assembly: nanostructures made easy}.  \jt{Advanced Materials}  \bvol{11}~(7),  \pg{579--585}.

\bibitem[Cheng {\em et~al.\/}(2016)Cheng, Zhang, Chen \& Hu]{cheng2016}
{\sc \au{Cheng, W.}, \au{Zhang, W.}, \au{Chen, H.} \& \au{Hu, L.}} \yr{2016}  \at{Spray cooling and flash evaporation cooling: The current development and application}.  \jt{Renewable and Sustainable Energy Reviews}  \bvol{55},  \pg{614--628}.

\bibitem[Chong {\em et~al.\/}(2020)Chong, Li, Ng, Verzicco \& Lohse]{chong2020}
{\sc \au{Chong, K.~L.}, \au{Li, Y.}, \au{Ng, C.~S.}, \au{Verzicco, R.} \& \au{Lohse, D.}} \yr{2020}  \at{Convection-dominated dissolution for single and multiple immersed sessile droplets}.  \jt{Journal of Fluid Mechanics}  \bvol{892},  \pg{A21}.

\bibitem[Cira {\em et~al.\/}(2015)Cira, Benusiglio \& Prakash]{cira2015}
{\sc \au{Cira, N.~J.}, \au{Benusiglio, A.} \& \au{Prakash, M.}} \yr{2015}  \at{Vapour-mediated sensing and motility in two-component droplets}.  \jt{Nature}  \bvol{519}~(7544),  \pg{446--450}.

\bibitem[Cussler(2009)]{cussler2009}
{\sc \au{Cussler, E.~L.}} \yr{2009} {\em Diffusion: mass transfer in fluid systems\/}.  \publ{Cambridge University Press}.

\bibitem[De~Gennes(1985)]{degennes1985}
{\sc \au{De~Gennes, P.~G.}} \yr{1985}  \at{Wetting: statics and dynamics}.  \jt{Reviews of Modern Physics}  \bvol{57}~(3),  \pg{827}.

\bibitem[Deegan {\em et~al.\/}(1997)Deegan, Bakajin, Dupont, Huber, Nagel \& Witten]{deegan1997}
{\sc \au{Deegan, R.~D.}, \au{Bakajin, O.}, \au{Dupont, T.~F.}, \au{Huber, G.}, \au{Nagel, S.~R.} \& \au{Witten, T.~A.}} \yr{1997}  \at{Capillary flow as the cause of ring stains from dried liquid drops}.  \jt{Nature}  \bvol{389}~(6653),  \pg{827--829}.

\bibitem[Deegan {\em et~al.\/}(2000)Deegan, Bakajin, Dupont, Huber, Nagel \& Witten]{deegan2000}
{\sc \au{Deegan, R.~D.}, \au{Bakajin, O.}, \au{Dupont, T.~F.}, \au{Huber, G.}, \au{Nagel, S.~R.} \& \au{Witten, T.~A.}} \yr{2000}  \at{Contact line deposits in an evaporating drop}.  \jt{Physical Review E}  \bvol{62}~(1),  \pg{756}.

\bibitem[Diddens(2017)]{diddens2017jcs}
{\sc \au{Diddens, C}} \yr{2017}  \at{Detailed finite element method modeling of evaporating multi-component droplets}.  \jt{Journal of Computational Physics}  \bvol{340},  \pg{670--687}.

\bibitem[Diddens {\em et~al.\/}(2024)Diddens, Dekker \& Lohse]{diddens2024}
{\sc \au{Diddens, Christian}, \au{Dekker, Pim~J} \& \au{Lohse, Detlef}} \yr{2024}  \at{Non-monotonic surface tension leads to spontaneous symmetry breaking in a binary evaporating drop}.  \jt{arXiv preprint arXiv:2402.17452} .

\bibitem[Diddens {\em et~al.\/}(2021)Diddens, Li \& Lohse]{diddens2021}
{\sc \au{Diddens, C.}, \au{Li, Y.} \& \au{Lohse, D.}} \yr{2021}  \at{Competing {{Marangoni}} and {{Rayleigh}} convection in evaporating binary droplets}.  \jt{Journal of Fluid Mechanics}  \bvol{914},  \pg{A23}.

\bibitem[Diddens {\em et~al.\/}(2017)Diddens, Tan, Lv, Versluis, Kuerten, Zhang \& Lohse]{diddens2017}
{\sc \au{Diddens, C}, \au{Tan, H.}, \au{Lv, P.}, \au{Versluis, M.}, \au{Kuerten, J. G.~M.}, \au{Zhang, X.} \& \au{Lohse, D.}} \yr{2017}  \at{Evaporating pure, binary and ternary droplets: thermal effects and axial symmetry breaking}.  \jt{Journal of Fluid Mechanics}  \bvol{823},  \pg{470--497}.

\bibitem[Dunn {\em et~al.\/}(2009)Dunn, Wilson, Duffy, David \& Sefiane]{dunn2009}
{\sc \au{Dunn, G.~J.}, \au{Wilson, S.~K.}, \au{Duffy, B.~R.}, \au{David, S.} \& \au{Sefiane, K.}} \yr{2009}  \at{The strong influence of substrate conductivity on droplet evaporation}.  \jt{Journal of Fluid Mechanics}  \bvol{623},  \pg{329--351}.

\bibitem[Edwards {\em et~al.\/}(2018)Edwards, Atkinson, Cheung, Liang, Fairhurst \& Ouali]{edwards2018}
{\sc \au{Edwards, A. M.~J.}, \au{Atkinson, P.~S.}, \au{Cheung, C.~S.}, \au{Liang, H.}, \au{Fairhurst, D.~J.} \& \au{Ouali, F.~F.}} \yr{2018}  \at{Density-driven flows in evaporating binary liquid droplets}.  \jt{Physical Review Letters}  \bvol{121}~(18),  \pg{184501}.

\bibitem[Edwards {\em et~al.\/}(2021)Edwards, Cater, Kilbride, Le~Minter, Brown, Fairhurst \& Ouali]{edwards2021}
{\sc \au{Edwards, A. M.~J.}, \au{Cater, J.}, \au{Kilbride, J.~J.}, \au{Le~Minter, P.}, \au{Brown, C.~V.}, \au{Fairhurst, D.~J.} \& \au{Ouali, F.~F.}} \yr{2021}  \at{Interferometric measurement of co-operative evaporation in 2d droplet arrays}.  \jt{Applied Physics Letters}  \bvol{119}~(15).

\bibitem[Fabrikant(1985)]{fabrikant1985}
{\sc \au{Fabrikant, V.~I.}} \yr{1985}  \at{On the potential flow through membranes}.  \jt{Zeitschrift f{\"u}r angewandte Mathematik und Physik ZAMP}  \bvol{36}~(4),  \pg{616--623}.

\bibitem[van Gaalen {\em et~al.\/}(2022)van Gaalen, Wijshoff, Kuerten \& Diddens]{gaalen2022}
{\sc \au{van Gaalen, R.~T.}, \au{Wijshoff, H. M.~A.}, \au{Kuerten, J. G.~M.} \& \au{Diddens, C}} \yr{2022}  \at{Competition between thermal and surfactant-induced marangoni flow in evaporating sessile droplets}.  \jt{Journal of Colloid and Interface Science}  \bvol{622},  \pg{892--903}.

\bibitem[Gelderblom {\em et~al.\/}(2022)Gelderblom, Diddens \& Marin]{gelderblom2022}
{\sc \au{Gelderblom, H.}, \au{Diddens, C.} \& \au{Marin, A.}} \yr{2022}  \at{Evaporation-driven liquid flow in sessile droplets}.  \jt{Soft Matter}  \bvol{18}~(45),  \pg{8535--8553}.

\bibitem[Gimenes {\em et~al.\/}(2013)Gimenes, Zhu, Raetano \& Oliveira]{gimenes2013}
{\sc \au{Gimenes, M.~J.}, \au{Zhu, H.}, \au{Raetano, C.~G.} \& \au{Oliveira, R.~B.}} \yr{2013}  \at{Dispersion and evaporation of droplets amended with adjuvants on soybeans}.  \jt{Crop Protection}  \bvol{44},  \pg{84--90}.

\bibitem[Hack {\em et~al.\/}(2021)Hack, Kwiecinski, Ram{\'\i}rez-Soto, Segers, Karpitschka, Kooij \& Snoeijer]{hack2021}
{\sc \au{Hack, M.~A.}, \au{Kwiecinski, W.}, \au{Ram{\'\i}rez-Soto, O.}, \au{Segers, T.}, \au{Karpitschka, S.}, \au{Kooij, E.~S.} \& \au{Snoeijer, J.~H.}} \yr{2021}  \at{Wetting of two-component drops: Marangoni contraction versus autophobing}.  \jt{Langmuir}  \bvol{37}~(12),  \pg{3605--3611}.

\bibitem[Hatte {\em et~al.\/}(2019)Hatte, Pandey, Pandey, Chakraborty \& Basu]{hatte2019}
{\sc \au{Hatte, S.}, \au{Pandey, K.}, \au{Pandey, K.}, \au{Chakraborty, S.} \& \au{Basu, S.}} \yr{2019}  \at{Universal evaporation dynamics of ordered arrays of sessile droplets}.  \jt{Journal of Fluid Mechanics}  \bvol{866},  \pg{61--81}.

\bibitem[Hoath(2016)]{hoath2016}
{\sc \au{Hoath, S.~D.}} \yr{2016} {\em Fundamentals of inkjet printing: the science of inkjet and droplets\/}.  \publ{John Wiley \& Sons}.

\bibitem[Kant {\em et~al.\/}(2024)Kant, Souzy, Kim, van~der Meer \& Lohse]{kant2024}
{\sc \au{Kant, P.}, \au{Souzy, M.}, \au{Kim, N.}, \au{van~der Meer, D.} \& \au{Lohse, D.}} \yr{2024}  \at{Autothermotaxis of volatile drops}.  \jt{Physical Review Fluids}  \bvol{9}~(1),  \pg{L012001}.

\bibitem[Karpitschka {\em et~al.\/}(2017)Karpitschka, Liebig \& Riegler]{karpitschka2017}
{\sc \au{Karpitschka, S.}, \au{Liebig, F.} \& \au{Riegler, H.}} \yr{2017}  \at{Marangoni contraction of evaporating sessile droplets of binary mixtures}.  \jt{Langmuir}  \bvol{33}~(19),  \pg{4682--4687}.

\bibitem[Khilifi {\em et~al.\/}(2019)Khilifi, Foudhil, Fahem, Harmand \& Ben]{khilifi2019}
{\sc \au{Khilifi, D.}, \au{Foudhil, W.}, \au{Fahem, K.}, \au{Harmand, S.} \& \au{Ben, J.~S.}} \yr{2019}  \at{Study of the phenomenon of the interaction between sessile drops during evaporation}.  \jt{Thermal Science}  \bvol{23}~(2 Part B),  \pg{1105--1114}.

\bibitem[Laghezza {\em et~al.\/}(2016)Laghezza, Dietrich, Yeomans, Ledesma-Aguilar, Kooij, Zandvliet \& Lohse]{laghezza2016}
{\sc \au{Laghezza, G.}, \au{Dietrich, E.}, \au{Yeomans, J.~M.}, \au{Ledesma-Aguilar, R.}, \au{Kooij, E.~S.}, \au{Zandvliet, H. J.~W.} \& \au{Lohse, D.}} \yr{2016}  \at{Collective and convective effects compete in patterns of dissolving surface droplets}.  \jt{Soft Matter}  \bvol{12}~(26),  \pg{5787--5796}.

\bibitem[Larson(2014)]{larson2014}
{\sc \au{Larson, R.~G.}} \yr{2014}  \at{Transport and deposition patterns in drying sessile droplets}.  \jt{AIChE Journal}  \bvol{60}~(5),  \pg{1538--1571}.

\bibitem[Legros {\em et~al.\/}(2015)Legros, Gaponenko, Mialdun, Triller, Hammon, Bauer, K{\"o}hler \& Shevtsova]{legros2015}
{\sc \au{Legros, J.~C.}, \au{Gaponenko, Y.}, \au{Mialdun, A.}, \au{Triller, T.}, \au{Hammon, A.}, \au{Bauer, C.}, \au{K{\"o}hler, W.} \& \au{Shevtsova, V.}} \yr{2015}  \at{Investigation of fickian diffusion in the ternary mixtures of water--ethanol--triethylene glycol and its binary pairs}.  \jt{Physical Chemistry Chemical Physics}  \bvol{17}~(41),  \pg{27713--27725}.

\bibitem[Li {\em et~al.\/}(2018)Li, Lv, Diddens, Tan, Wijshoff, Versluis \& Lohse]{li2018}
{\sc \au{Li, Y.}, \au{Lv, P.}, \au{Diddens, C.}, \au{Tan, H.}, \au{Wijshoff, H.}, \au{Versluis, M.} \& \au{Lohse, D.}} \yr{2018}  \at{Evaporation-triggered segregation of sessile binary droplets}.  \jt{Physical Review Letters}  \bvol{120}~(22),  \pg{224501}.

\bibitem[Lohse(2022)]{lohse2022}
{\sc \au{Lohse, D.}} \yr{2022}  \at{Fundamental fluid dynamics challenges in inkjet printing}.  \jt{Annual Review of Fluid Mechanics}  \bvol{54},  \pg{349--382}.

\bibitem[Masoud {\em et~al.\/}(2021)Masoud, Howell \& Stone]{masoud2021}
{\sc \au{Masoud, H.}, \au{Howell, P.~D.} \& \au{Stone, H.~A.}} \yr{2021}  \at{Evaporation of multiple droplets}.  \jt{Journal of Fluid Mechanics}  \bvol{927},  \pg{R4}.

\bibitem[Nguyen {\em et~al.\/}(2012)Nguyen, Nguyen, Hampton, Xu, Huang \& Rudolph]{nguyen2012ces}
{\sc \au{Nguyen, T. A.~H.}, \au{Nguyen, A.~V.}, \au{Hampton, M.~A.}, \au{Xu, Z.~P.}, \au{Huang, L.} \& \au{Rudolph, V.}} \yr{2012}  \at{Theoretical and experimental analysis of droplet evaporation on solid surfaces}.  \jt{Chemical Engineering Science}  \bvol{69}~(1),  \pg{522--529}.

\bibitem[Pahlavan {\em et~al.\/}(2021)Pahlavan, Yang, Bain \& Stone]{pahlavan2021}
{\sc \au{Pahlavan, A.~A.}, \au{Yang, L.}, \au{Bain, C.~D.} \& \au{Stone, H.~A.}} \yr{2021}  \at{Evaporation of binary-mixture liquid droplets: the formation of picoliter pancakelike shapes}.  \jt{Physical Review Letters}  \bvol{127}~(2),  \pg{024501}.

\bibitem[Popov(2005)]{popov2005}
{\sc \au{Popov, Y.~O.}} \yr{2005}  \at{Evaporative deposition patterns: spatial dimensions of the deposit}.  \jt{Physical Review E}  \bvol{71}~(3),  \pg{036313}.

\bibitem[Ristenpart {\em et~al.\/}(2007)Ristenpart, Kim, Domingues, Wan \& Stone]{ristenpart2007}
{\sc \au{Ristenpart, W.~D.}, \au{Kim, P.~G.}, \au{Domingues, C.}, \au{Wan, J.} \& \au{Stone, H.~A.}} \yr{2007}  \at{Influence of substrate conductivity on circulation reversal in evaporating drops}.  \jt{Physical Review Letters}  \bvol{99}~(23),  \pg{234502}.

\bibitem[Rocha {\em et~al.\/}(2024)Rocha, Lederer, Dekker, Marin, Lohse \& Diddens]{rocha2024}
{\sc \au{Rocha, D.}, \au{Lederer, P.~L.}, \au{Dekker, P.~J.}, \au{Marin, A.}, \au{Lohse, D.} \& \au{Diddens, C.}} \yr{2024}  \at{Evaporating sessile droplets: solutal marangoni effects overwhelm thermal marangoni flow}.  \jt{arXiv preprint arXiv:2410.17071} .

\bibitem[Schofield {\em et~al.\/}(2020)Schofield, Wray, Pritchard \& Wilson]{schofield2020}
{\sc \au{Schofield, F. G.~H.}, \au{Wray, A.~W.}, \au{Pritchard, D.} \& \au{Wilson, S.~K.}} \yr{2020}  \at{The shielding effect extends the lifetimes of two-dimensional sessile droplets}.  \jt{Journal of Engineering Mathematics}  \bvol{120}~(1),  \pg{89--110}.

\bibitem[Shiri {\em et~al.\/}(2021)Shiri, Sinha, Baumgartner \& Cira]{shiri2021}
{\sc \au{Shiri, S.}, \au{Sinha, S.}, \au{Baumgartner, D.~A.} \& \au{Cira, N.~J.}} \yr{2021}  \at{Thermal marangoni flow impacts the shape of single component volatile droplets on thin, completely wetting substrates}.  \jt{Physical Review Letters}  \bvol{127}~(2),  \pg{024502}.

\bibitem[Silberzan {\em et~al.\/}(1991)Silberzan, Leger, Ausserre \& Benattar]{silberzan1991}
{\sc \au{Silberzan, P.}, \au{Leger, L.}, \au{Ausserre, D.} \& \au{Benattar, J.~J.}} \yr{1991}  \at{Silanation of silica surfaces. a new method of constructing pure or mixed monolayers}.  \jt{Langmuir}  \bvol{7}~(8),  \pg{1647--1651}.

\bibitem[Stauber {\em et~al.\/}(2014)Stauber, Wilson, Duffy \& Sefiane]{stauber2014}
{\sc \au{Stauber, J.~M.}, \au{Wilson, S.~K.}, \au{Duffy, B.~R.} \& \au{Sefiane, K.}} \yr{2014}  \at{On the lifetimes of evaporating droplets}.  \jt{Journal of Fluid Mechanics}  \bvol{744},  \pg{R2}.

\bibitem[Stauber {\em et~al.\/}(2015)Stauber, Wilson, Duffy \& Sefiane]{stauber2015}
{\sc \au{Stauber, J.~M.}, \au{Wilson, S.~K.}, \au{Duffy, B.~R.} \& \au{Sefiane, K.}} \yr{2015}  \at{On the lifetimes of evaporating droplets with related initial and receding contact angles}.  \jt{Physics of fluids}  \bvol{27}~(12).

\bibitem[Thayyil~Raju {\em et~al.\/}(2022)Thayyil~Raju, Diddens, Li, Marin, van~der Linden, Zhang \& Lohse]{raju2022}
{\sc \au{Thayyil~Raju, L.}, \au{Diddens, C.}, \au{Li, Y.}, \au{Marin, A.}, \au{van~der Linden, M.~N.}, \au{Zhang, X.} \& \au{Lohse, D.}} \yr{2022}  \at{Evaporation of a sessile colloidal water--glycerol droplet: Marangoni ring formation}.  \jt{Langmuir}  \bvol{38}~(39),  \pg{12082--12094}.

\bibitem[Wang {\em et~al.\/}(2024)Wang, Karapetsas, Valluri \& Inoue]{wang2024apl}
{\sc \au{Wang, Z.}, \au{Karapetsas, G.}, \au{Valluri, P.} \& \au{Inoue, C.}} \yr{2024}  \at{Flow structure near three phase contact line of low-contact-angle evaporating droplets}.  \jt{Applied Physics Letters}  \bvol{124}~(10).

\bibitem[Wang {\em et~al.\/}(2022)Wang, Orejon, Takata \& Sefiane]{wang2022}
{\sc \au{Wang, Z.}, \au{Orejon, D.}, \au{Takata, Y.} \& \au{Sefiane, K.}} \yr{2022}  \at{Wetting and evaporation of multicomponent droplets}.  \jt{Physics Reports}  \bvol{960},  \pg{1--37}.

\bibitem[Wijshoff(2018)]{wijshoff2018}
{\sc \au{Wijshoff, H.}} \yr{2018}  \at{Drop dynamics in the inkjet printing process}.  \jt{Current Opinion in Colloid \& Interface Science}  \bvol{36},  \pg{20--27}.

\bibitem[Wilson \& D'Ambrosio(2023)]{wilson2023}
{\sc \au{Wilson, S.~K.} \& \au{D'Ambrosio, H.}} \yr{2023}  \at{Evaporation of sessile droplets}.  \jt{Annual Review of Fluid Mechanics}  \bvol{55},  \pg{481--509}.

\bibitem[Wray {\em et~al.\/}(2020)Wray, Duffy \& Wilson]{wray2020}
{\sc \au{Wray, A.~W.}, \au{Duffy, B.~R.} \& \au{Wilson, S.~K.}} \yr{2020}  \at{Competitive evaporation of multiple sessile droplets}.  \jt{Journal of Fluid Mechanics}  \bvol{884},  \pg{A45}.

\bibitem[Wray {\em et~al.\/}(2021)Wray, Wray, Duffy \& Wilson]{wray2021}
{\sc \au{Wray, A.~W.}, \au{Wray, P.~S.}, \au{Duffy, B.~R.} \& \au{Wilson, S.~K.}} \yr{2021}  \at{Contact-line deposits from multiple evaporating droplets}.  \jt{Physical Review Fluids}  \bvol{6}~(7),  \pg{073604}.

\bibitem[Yu {\em et~al.\/}(2009)Yu, Zhu, Frantz, Reding, Chan \& Ozkan]{yu2009}
{\sc \au{Yu, Y.}, \au{Zhu, H.}, \au{Frantz, J.~M.}, \au{Reding, M.~E.}, \au{Chan, K.~C.} \& \au{Ozkan, H.~E.}} \yr{2009}  \at{Evaporation and coverage area of pesticide droplets on hairy and waxy leaves}.  \jt{Biosyst. Eng.}  \bvol{104}~(3),  \pg{324--334}.

\bibitem[Zang {\em et~al.\/}(2019)Zang, Tarafdar, Tarasevich, Choudhury \& Dutta]{zang2019}
{\sc \au{Zang, D.}, \au{Tarafdar, S.}, \au{Tarasevich, Y.~Y.}, \au{Choudhury, M.~D.} \& \au{Dutta, T.}} \yr{2019}  \at{Evaporation of a droplet: From physics to applications}.  \jt{Physics Reports}  \bvol{804},  \pg{1--56}.

\bibitem[Zhang {\em et~al.\/}(2015)Zhang, Wang, Guo, Zamora, Ying, Lin, Wang, Hu \& Wang]{zhang2015}
{\sc \au{Zhang, R.}, \au{Wang, G.}, \au{Guo, S.}, \au{Zamora, M.~L.}, \au{Ying, Q.}, \au{Lin, Y.}, \au{Wang, W.}, \au{Hu, M.} \& \au{Wang, Y.}} \yr{2015}  \at{Formation of urban fine particulate matter}.  \jt{Chemical Reviews}  \bvol{115}~(10),  \pg{3803--3855}.

\bibitem[Zuend {\em et~al.\/}(2008)Zuend, Marcolli, Luo \& Peter]{zuend2008}
{\sc \au{Zuend, A.}, \au{Marcolli, C.}, \au{Luo, B.~P.} \& \au{Peter, T.}} \yr{2008}  \at{A thermodynamic model of mixed organic-inorganic aerosols to predict activity coefficients}.  \jt{Atmospheric Chemistry and Physics}  \bvol{8}~(16),  \pg{4559--4593}.

\end{thebibliography}
\end{document}